\documentclass[journal]{IEEEtran}

\usepackage{cite}                
\usepackage{amsmath,amssymb}     
\usepackage{graphicx}            
\usepackage{textcomp}            
\usepackage{xcolor}              
\usepackage{pifont}              
\usepackage{booktabs}            
\usepackage{multirow}            
\usepackage{url}                 
\usepackage{hyperref}            
\usepackage{caption}             
\usepackage{longtable}           
\usepackage{float}               
\usepackage{tabularx}   
\usepackage{pifont}
\usepackage{pgfplots}
\usepackage{subcaption}
\usetikzlibrary{positioning, arrows.meta, calc, shapes.geometric}
\usepackage{pgfplots}
\usepackage{algorithm}      
\usepackage{algpseudocode} 
\usepackage{tikz}          
\usepackage{pgfplots}      
\pgfplotsset{compat=1.18}  

\pgfplotsset{compat=1.18}
\usetikzlibrary{positioning, arrows.meta, calc, shapes.geometric}
\usetikzlibrary{shapes.multipart, arrows.meta, positioning}

\pgfplotsset{compat=1.17}

\newcommand{\cmark}{\ding{51}} 
\newcommand{\xmark}{\ding{55}} 

\begin{document}
	
	\title{FedFog: Resource-Aware Federated Learning in Edge and Fog Networks}

\author{Somayeh Sobati-M.
	
	\thanks{Somayeh Sobati-M. is with the Department
		of Electrical and Computer Engineering, Hakim Sabzevari University, Sabzevar, Iran. Somayeh Sobati-M.  is also with the Department of Computer Engineering,
		Ferdowsi University of Mashhad, Mashhad, Iran, e-mail:s.sobati@hsu.ac.ir}
	\thanks{Manuscript received April 19, 2024; revised January 11, 2024.}}

	\maketitle
	
	\begin{abstract}
		As edge and fog computing become central to modern distributed systems, there’s growing interest in combining serverless architectures with privacy-preserving machine learning techniques like federated learning (FL). However, current simulation tools fail to capture this integration effectively. In this paper, we introduce FedFog, a simulation framework that extends the FogFaaS environment to support FL-aware serverless execution across edge–fog infrastructures. FedFog incorporates an adaptive FL scheduler, privacy-respecting data flow, and resource-aware orchestration to emulate realistic, dynamic conditions in IoT-driven scenarios. Through extensive simulations on benchmark datasets, we demonstrate that FedFog accelerates model convergence, reduces latency, and improves energy efficiency compared to conventional FL or FaaS setups—making it a valuable tool for researchers exploring scalable, intelligent edge systems.
	\end{abstract}
	

\section{Introduction}

The rapid growth of Internet of Things (IoT) applications has led to an increased demand for computing paradigms that process data near its source. Edge and fog computing offer promising solutions by distributing computation closer to end devices, reducing both latency and bandwidth usage~\cite{9134408}. Simulation frameworks such as iFogSim~\cite{gupta2017ifogsim} and iFogSim2~\cite{mahmud2020ifogsim2} have proven valuable for evaluating resource management and scheduling in these distributed environments.

However, these frameworks do not support serverless computing models, which are increasingly prevalent in cloud-native and event-driven architectures. Serverless paradigms, or Function-as-a-Service (FaaS), offer dynamic scalability and are well-suited for edge computing scenarios that involve sensitive or distributed data. While FL libraries such as Flower~\cite{beutel2022flower} enable flexible experimentation with learning algorithms, they do not simulate the underlying constraints of edge environments—such as network unreliability, limited compute, or energy-aware scheduling~\cite{10210010}.

Moreover, existing tools lack orchestration support for FL in serverless environments. This omission is critical, as real-world FL deployments must address client heterogeneity, data drift, and dynamic participation, all while operating under resource constraints.

This divergence between theoretical algorithmic capabilities and practical implementation requirements reveals a critical research void that must be addressed. In operational environments, FL systems encounter four interconnected challenges that collectively complicate deployment. First, the inherently dynamic nature of client participation stems from device mobility patterns and finite energy budgets, creating unpredictable availability windows. Second, the statistical distribution of data varies significantly across edge nodes, violating the independent and identically distributed (IID) assumptions common to many machine learning approaches. Third, the serverless execution model introduces unpredictable cold start latencies during function initialization, which can disrupt training timelines. Finally, practitioners must navigate fundamental tensions between model accuracy objectives and system efficiency metrics, requiring careful optimization across multiple competing dimensions. These operational realities remain largely unaddressed by current simulation tools, limiting their utility for real-world deployment planning.\par
In this paper, we introduce \textbf{FedFog}—a modular simulation framework that integrates FL with serverless execution atop the iFogSim environment. FedFog supports dynamic function orchestration, resource-aware scheduling, node health monitoring, and data drift simulation. It enables detailed evaluation of FL systems under realistic edge conditions, including client dropout, cold starts, and non-IID data.

Our work is motivated by the practical challenges faced when modeling FL in constrained and heterogeneous environments. FedFog aims to fill this simulation gap and serve both as a research tool and a testbed for robust and efficient FL algorithm design.

\textbf{Our key contributions are:}
\begin{itemize}
	\item Design and implementation of FedFog, a serverless FL simulator built on iFogSim.
	\item Support for cold start simulation, client heterogeneity, energy constraints, and data drift.
	\item Comprehensive evaluation against FaaS and centralized learning baselines across multiple metrics (accuracy, latency, energy).
\end{itemize}
The remainder of this paper is organized as follows. Section~\ref{Sec.RelatedWorks} surveys the most relevant simulation tools and federated learning frameworks, identifying the limitations that motivate our work. Section~\ref{Sec.ProposedMethod} introduces the architecture and design principles of the proposed FedFog framework, including its adaptive scheduling and cold start modeling. Section~\ref{Sec.ExperimentalEvaluation} presents the experimental setup, evaluation metrics, and comparative results under various conditions, including adversarial scenarios and real-world deployment. Section~\ref{Sec.TheoreticalComplexityAnalysis} provides a theoretical analysis of FedFog’s complexity and trade-offs, supported by sensitivity and scalability experiments. Finally, Section~\ref{Sec.Conclusion} concludes the paper and outlines future directions, including enhanced privacy mechanisms and trust-aware scheduling.

\section{Related Work}\label{Sec.RelatedWorks}

This section systematically reviews prior research in three key areas foundational to our work. First, we analyze \textit{iFogSim} and \textit{iFogSim2} as seminal simulators for fog/edge environments, focusing on their resource modeling capabilities and limitations in handling machine learning workloads. Second, we examine advancements in \textit{Serverless Computing}, particularly \textit{FogFaaS} extensions, that enable efficient function orchestration across edge-fog hierarchies. Finally, we survey \textit{FL Frameworks}, emphasizing their adaptation to distributed infrastructure constraints and privacy requirements. By synthesizing these strands of research, we identify critical gaps—especially in simulating FL-serverless integration—that motivate our FedFog framework.

Serverless computing, or Function-as-a-Service (FaaS), has emerged as a paradigm for scalable and event-driven applications. OpenFaaS \cite{williams2017openfaas} is an open-source FaaS platform that enables developers to deploy functions over Docker and Kubernetes. While effective in production environments, OpenFaaS lacks simulation capabilities and is unsuitable for evaluating large-scale edge scenarios under controlled conditions.
The iFogSim toolkit \cite{gupta2017ifogsim} is a widely used simulation framework for modeling and evaluating resource management strategies in edge and fog computing environments. It supports latency-aware resource allocation, hierarchical topologies, and basic mobility. However, iFogSim lacks built-in support for serverless computing paradigms and machine learning workflows. To address some of these gaps, iFogSim2 \cite{mahmud2020ifogsim2} was introduced, which enhances mobility modeling and introduces microservice management and clustering capabilities. Despite these improvements, neither framework supports serverless computing for AI-driven edge applications.

To fill this gap, Ghaseminya et al. proposed FogFaaS \cite{ghaseminya2025fogfaas}, an extension of iFogSim that simulates serverless computing in fog environments. FogFaaS introduces components to model cold starts, dynamic resource scaling, and function invocation behavior \cite{li2025pfedkd}. However, it does not support AI workloads, especially FL, which is increasingly relevant in edge scenarios where privacy and data locality are essential.

FL allows decentralized model training across multiple devices while preserving data privacy~\cite{10542323}. The foundational work by McMahan et al. \cite{mcmahan2017communication} introduced the Federated Averaging (FedAvg) algorithm, which has become the basis for many FL implementations. Flower \cite{beutel2022flower} is a flexible FL framework that supports different ML backends and training configurations. Despite its flexibility, Flower does not offer simulation of edge resource constraints or integration with FaaS execution models.

\subsection{Gap and Motivation}

While several simulation tools support edge/fog environments and others focus on FL experimentation, none of the existing platforms provide a unified simulation of FL-aware serverless computing (See Table~\ref{tab:comparison}). There is a critical need for a framework that combines the dynamic scalability of FaaS with the intelligence and privacy-preserving nature of FL in resource-constrained edge environments. This motivates the development of \textbf{FedFog}, a federated-learning-aware serverless simulation framework that extends FogFaaS by integrating FL task scheduling, model aggregation, and resource-aware orchestration within iFogSim.

\begin{table*}[ht]
	\centering
	\caption{Comparison of Existing Frameworks with the Proposed \textbf{FedFog} Simulator}
	\label{tab:comparison}
	\begin{tabular}{|p{4.3cm}|c|c|c|c|c|c|}
		\hline
		\textbf{Feature / Framework} & \textbf{iFogSim \cite{gupta2017ifogsim}} & \textbf{FogFaaS \cite{ghaseminya2025fogfaas}} & \textbf{OpenFaaS \cite{williams2017openfaas}} & \textbf{Flower \cite{beutel2022flower}} & \textbf{iFogSim2 \cite{mahmud2020ifogsim2}} & \textbf{FedFog (Proposed)} \\
		\hline
		FaaS Support                & \xmark           & \cmark           & \cmark            & \xmark          & \xmark            & \cmark \\
		\hline
		Simulation Capability       & \cmark           & \cmark           & \xmark            & \xmark          & \cmark            & \cmark \\
		\hline
		Federated Learning          & \xmark           & \xmark           & \xmark            & \cmark          & \xmark            & \cmark \\
		\hline
		Edge Resource Modeling      & \cmark           & \cmark           & \xmark            & \xmark          & \cmark            & \cmark \\
		\hline
		Cold Start Modeling         & \xmark           & \cmark           & \xmark            & \xmark          & \xmark            & \cmark \\
		\hline
		FL-Aware Scheduling         & \xmark           & \xmark           & \xmark            & \xmark          & \xmark            & \cmark \\
		\hline
		Dynamic Function Scaling    & \xmark           & \cmark           & \cmark            & \xmark          & \xmark            & \cmark \\
		\hline
		Privacy-Preserving ML       & \xmark           & \xmark           & \xmark            & \cmark          & \xmark            & \cmark \\
		\hline
	\end{tabular}
\end{table*}

\section{The Proposed Method}\label{Sec.ProposedMethod}

\begin{figure*}[htbp]
	\centering
	\includegraphics[width=1.0\linewidth]{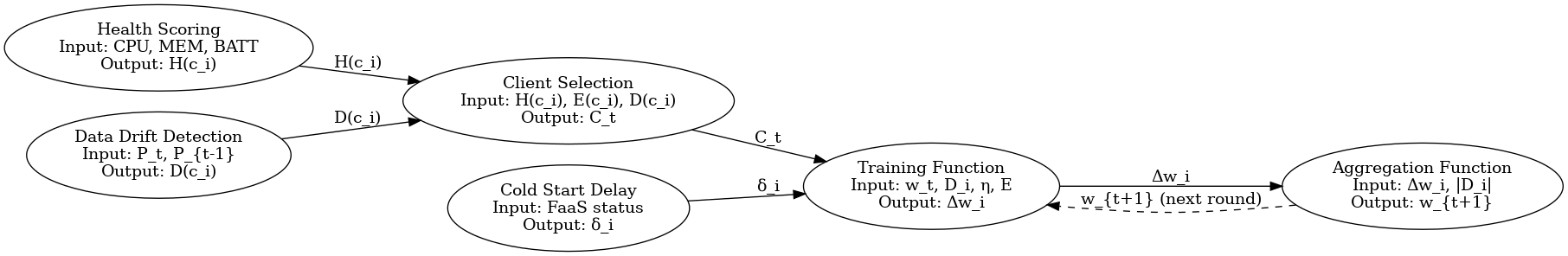}
	\caption{Functional flow diagram of the FedFog framework.}
	\label{fig:fedfog-flow}
\end{figure*}

This section follows the dataflow of the FedFog framework as illustrated in Figure~\ref{fig:fedfog-flow}, detailing the input-output behavior of each core module and how they interconnect to support privacy-aware, adaptive FL at the edge. Initially, health scores and data drift metrics are computed independently and then used as input criteria for the client selection function. Clients that satisfy the selection thresholds proceed to participate in local model training, which is executed in a serverless environment~\cite{10542323}. Each client’s model update is then forwarded to the aggregation function, where a new global model is computed. Additionally, cold start latency affects the execution time of training functions and is considered during simulation. The resulting global model is then distributed to clients for the next FL round, completing the cycle.

\subsection{Health Scoring Function}

Each edge client \( c_i \) is evaluated based on local system resources, including CPU availability, memory availability, and battery level. These raw metrics are normalized and combined into a single scalar value representing the client’s health.

\begin{equation}\label{Eq.HealthScore}
	H(c_i) = \alpha_1 \cdot \text{CPU}_i + \alpha_2 \cdot \text{MEM}_i + \alpha_3 \cdot \text{BATT}_i
\end{equation}

Here, \( \alpha_1 + \alpha_2 + \alpha_3 = 1 \) are predefined weights controlling the relative contribution of each metric. The output \( H(c_i) \) is a normalized score used to assess the suitability of client \( c_i \) for training.

\subsection{Data Drift Detection Function}

To ensure data stability, the FedFog framework monitors changes in the local data distribution across rounds. For client \( c_i \), a drift score is computed using Kullback–Leibler (KL) divergence:

\begin{equation}\label{Eq.DriftDetection}
	D(c_i) = \text{KL}(P_t(D_i) \parallel P_{t-1}(D_i))
\end{equation}

Where \( P_t(D_i) \) and \( P_{t-1}(D_i) \) are the empirical class or feature distributions of client \( i \)’s dataset in rounds \( t \) and \( t-1 \), respectively. A higher value of \( D(c_i) \) implies significant change in local data.

\subsection{Client Selection Function}

The FL scheduler selects clients to participate in round \( t \) using three criteria: health score, energy level, and data drift. The selection function is defined as:

\begin{equation}
	C_t = \left\{ c_i \in \mathcal{C} \mid H(c_i) > \theta_h \ \land \ E(c_i) > \theta_e \ \land \ D(c_i) < \theta_d \right\}\\
\end{equation}

Where \( \mathcal{C} \) is the set of all registered clients, and \( \theta_h \), \( \theta_e \), and \( \theta_d \) are threshold values. The output \( C_t \) is the set of selected clients for this training round.

\subsection{Cold Start Delay Function}

As training is executed using serverless computing, each function invocation may experience latency based on whether the execution container is already active. The delay function is modeled as:

\begin{equation}
	\delta_i =
	\begin{cases}
		\delta_{\text{cold}}, & \text{if first-time invocation} \\
		\delta_{\text{warm}}, & \text{otherwise}
	\end{cases}
\end{equation}

This delay \( \delta_i \) impacts the timing and responsiveness of the system but does not influence model correctness.

\subsection{Training Function (Serverless)}

Each selected client \( c_i \in C_t \) trains a local model update using its private dataset \( D_i \) and the shared global model \( w_t \). Training is executed in a stateless containerized function:

\begin{equation}
	\Delta w_i = \texttt{Train}(w_t, D_i, \eta, E)
\end{equation}

Where \( \eta \) is the learning rate, and \( E \) is the number of local epochs. The output \( \Delta w_i \) is the local model update returned to the fog node.

\subsection{Aggregation Function}

The fog node receives updates \( \Delta w_i \) from all selected clients and computes the new global model using Federated Averaging:

\begin{equation}
	w_{t+1} = \sum_{i \in C_t} \frac{|D_i|}{\sum_{j \in C_t} |D_j|} \cdot \Delta w_i
\end{equation}

Clients with larger local datasets contribute proportionally more to the global update. The output \( w_{t+1} \) is then distributed to clients in the next round, closing the feedback loop.

\subsection{Scheduler Utility Function}
\label{subsec:utility}
To prioritize clients for training, FedFog computes a \textbf{utility score} for each client $c_i$ based on their health, energy, and drift metrics. The utility function combines these factors into a single scalar value, enabling ranked client selection:

\begin{equation}
	U(c_i) = \beta_1 \cdot H(c_i) + \beta_2 \cdot E(c_i) - \beta_3 \cdot D(c_i)
	\label{eq:utility}
\end{equation}

\noindent where:
\begin{itemize}
	\item $H(c_i)$: Health score (Eq.~\ref{Eq.HealthScore})
	\item $E(c_i)$: Normalized energy level (battery/CPU availability)
	\item $D(c_i)$: Data drift score (Eq.~\ref{Eq.DriftDetection})
	\item $\beta_1, \beta_2, \beta_3$: Tunable weights ($\beta_1 + \beta_2 + \beta_3 = 1$)
\end{itemize}

\noindent \textbf{Interpretation}: Higher values of $H(c_i)$ and $E(c_i)$ increase the utility score, favoring clients with stable resource availability and sufficient energy reserves. Conversely, higher $D(c_i)$ values reduce utility, penalizing clients exhibiting significant data drift to maintain model consistency.

\subsection*{Comprehensive Numerical Example: Client Selection to Aggregation}

Consider a scenario with three clients \( c_1, c_2, c_3 \in \mathcal{C} \) and the following conditions:

\paragraph{Thresholds for selection:}
\[
\theta_h = 0.6, \quad \theta_e = 0.5, \quad \theta_d = 0.1
\]

\paragraph{Client attributes:}
\[
\begin{array}{|c|c|c|c|c|c|}
	\hline
	\text{Client} & \text{CPU} & \text{MEM} & \text{BATT} & E(c_i) & D(c_i) \\
	\hline
	c_1 & 0.8 & 0.6 & 0.5 & 0.7 & 0.05 \\
	c_2 & 0.4 & 0.5 & 0.4 & 0.6 & 0.12 \\
	c_3 & 0.9 & 0.7 & 0.8 & 0.9 & 0.02 \\
	\hline
\end{array}
\]

\paragraph{Step 1: Compute Health Scores (using \( \alpha_1 = 0.4, \alpha_2 = 0.3, \alpha_3 = 0.3 \))}

\[
H(c_1) = 0.4 \cdot 0.8 + 0.3 \cdot 0.6 + 0.3 \cdot 0.5 = 0.32 + 0.18 + 0.15 = 0.65
\]
\[
H(c_2) = 0.4 \cdot 0.4 + 0.3 \cdot 0.5 + 0.3 \cdot 0.4 = 0.16 + 0.15 + 0.12 = 0.43
\]
\[
H(c_3) = 0.4 \cdot 0.9 + 0.3 \cdot 0.7 + 0.3 \cdot 0.8 = 0.36 + 0.21 + 0.24 = 0.81
\]

\paragraph{Step 2: Apply Client Selection Function}

We include clients that satisfy: \( H(c_i) > \theta_h \), \( E(c_i) > \theta_e \), \( D(c_i) < \theta_d \)\\
- \( c_1 \): \( H=0.65 > 0.6, E=0.7 > 0.5, D=0.05 < 0.1 \Rightarrow \) selected\\
- \( c_2 \): \( H=0.43 < 0.6 \Rightarrow \) rejected\\
- \( c_3 \): \( H=0.81 > 0.6, E=0.9 > 0.5, D=0.02 < 0.1 \Rightarrow \) selected

\[
C_t = \{c_1, c_3\}
\]

\paragraph{Step 3: Local Training Outputs}

Assume global model \( w_t \) is trained for \( E = 3 \) epochs with learning rate \( \eta = 0.01 \). Each client returns updates:

\[
\Delta w_1 = [0.2, -0.1], \quad \Delta w_3 = [0.5, 0.0]
\]

\paragraph{Step 4: Aggregate Updates via FedAvg}

The fog node aggregates the updates from the selected clients using the Federated Averaging (FedAvg) algorithm. Given client dataset sizes \( |D_1| = 100 \) and \( |D_3| = 300 \), the weighted average is computed as follows:

\begin{align*}
	w_{t+1} &= \frac{100}{400} \cdot \Delta w_1 + \frac{300}{400} \cdot \Delta w_3 \\
	&= 0.25 \cdot [0.2, -0.1] + 0.75 \cdot [0.5, 0.0] \\
	&= [0.05, -0.025] + [0.375, 0.0] \\
	&= [0.425, -0.025]
\end{align*}

\paragraph{Step 6: Cold Start Delays}

Assume \( \delta_{\text{cold}} = 2000 \text{ms} \), \( \delta_{\text{warm}} = 200 \text{ms} \)

- If \( c_1 \) is invoked for the first time: \( \delta_1 = 2000 \)
- If \( c_3 \) was previously used: \( \delta_3 = 200 \)

\paragraph{Step 7: Scheduler Utility Scores}

Use: \( \beta_1 = 0.4, \beta_2 = 0.4, \beta_3 = 0.2 \)

\[
U(c_1) = 0.4 \cdot 0.65 + 0.4 \cdot 0.7 - 0.2 \cdot 0.05 = 0.26 + 0.28 - 0.01 = 0.53
\]
\[
U(c_3) = 0.4 \cdot 0.81 + 0.4 \cdot 0.9 - 0.2 \cdot 0.02 = 0.324 + 0.36 - 0.004 = 0.68
\]

So \( c_3 \) is the highest priority client.

\paragraph{Final Outcome:}
- Selected clients for training: \( \{c_1, c_3\} \)
- Aggregated model update: \( w_{t+1} = [0.425, -0.025] \)
- Scheduling order: \( c_3 \) has highest utility

\subsection{Trade-off Formalization}
\label{subsec:tradeoff}

FedFog is designed to intelligently navigate the inherent trade-offs between three critical objectives in federated edge environments: model accuracy, system latency, and energy efficiency. These objectives often conflict in practice, meaning that improving one may come at the cost of another. Therefore, the system must make informed scheduling decisions to achieve a desirable balance.

To formalize this challenge, we frame it as a constrained optimization problem. Let $\mathcal{C}_t$ denote the set of selected clients in round $t$, and let $w_T$ represent the global model after $T$ training rounds. Our goal is to maximize the model's predictive accuracy while keeping system latency and energy usage within acceptable limits. This can be expressed as:

\begin{equation}
	\begin{aligned}
		& \underset{\mathcal{C}_t}{\text{maximize}} 
		& & \text{Accuracy}(w_T) \\
		& \text{subject to}
		& & \text{Latency}(\delta_i, \mathcal{C}_t) \leq \tau_{\text{max}}, \\
		& & & \text{Energy}(E_i, \mathcal{C}_t) \leq \epsilon_{\text{max}}, \\
		& & & \mathcal{C}_t = \{c_i \mid H(c_i) > \theta_h, E(c_i) > \theta_e, D(c_i) < \theta_d\}\\
	\end{aligned}
\end{equation}

Here, $\tau_{\text{max}}$ and $\epsilon_{\text{max}}$ are user-defined thresholds specifying the maximum tolerable latency and energy usage per round. Clients are selected only if they meet the system health, energy, and data drift thresholds.

This optimization process gives rise to a Pareto frontier, where different client selection strategies lead to trade-offs among the competing objectives. For example, increasing the number of participating clients, $|\mathcal{C}_t|$, generally improves model accuracy by incorporating more diverse data. However, it also increases round latency because of slower devices, commonly referred to as stragglers. The latency per round $\delta_i$ typically scales with $\mathcal{O}(|\mathcal{C}_t|)$, so FedFog incorporates utility scores $U(c_i)$ to prioritize clients that offer the best trade-off between learning value and computational overhead.

Energy consumption presents a similar challenge. As more clients participate, total energy usage increases linearly—formally approximated by $E_{\text{total}} \approx \sum_{i \in \mathcal{C}_t} E_i$. To manage this, FedFog's scheduler uses the energy threshold $\theta_e$ to restrict participation to devices with sufficient battery levels, thereby maintaining energy sustainability without compromising learning effectiveness.

By modeling these trade-offs explicitly, FedFog adapts client selection dynamically based on resource availability, client utility, and system constraints. This design enables scalable, real-world deployments of FL on heterogeneous edge-fog infrastructures.

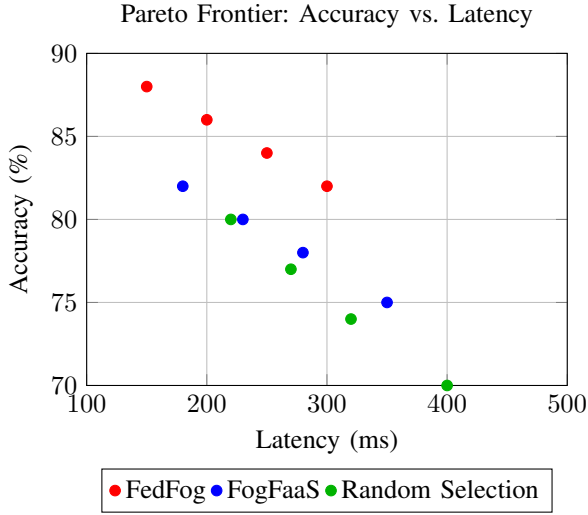
\begin{figure}[t]
	\centering
	\begin{tikzpicture}
		\begin{axis}[
			width=0.9\linewidth,
			height=6cm,
			xlabel={Latency (ms)},
			ylabel={Accuracy (\%)},
			legend style={
				at={(0.5,-0.25)},
				anchor=north,
				legend columns=3
			},
			grid=major,
			title={Pareto Frontier: Accuracy vs. Latency},
			xmin=100, xmax=500,
			ymin=70, ymax=90,
			axis background/.style={fill=white},
			]
			
			\addplot[
			only marks,
			mark=*,
			color=red,
			mark size=2pt
			] coordinates {
				(150, 88) (200, 86) (250, 84) (300, 82)
			};
			
			\addplot[
			only marks,
			mark=*,
			color=blue,
			mark size=2pt
			] coordinates {
				(180, 82) (230, 80) (280, 78) (350, 75)
			};
			
			\addplot[
			only marks,
			mark=*,
			color=green!70!black,
			mark size=2pt
			] coordinates {
				(220, 80) (270, 77) (320, 74) (400, 70)
			};
			
			\legend{FedFog, FogFaaS, Random Selection}
		\end{axis}
	\end{tikzpicture}
	\caption{Pareto frontier comparing FedFog’s accuracy-latency trade-off against baselines.}
	\label{fig:pareto}
\end{figure}

\vspace{1mm}
To illustrate these trade-offs visually, Figure~\ref{fig:pareto} presents a Pareto frontier showing the relationship between latency and accuracy across three different strategies: FedFog, FogFaaS, and Random Client Selection. Each point in the plot corresponds to a specific round configuration under increasing client load. 

The red points represent FedFog, which consistently achieves higher accuracy at lower latency compared to the other methods. This demonstrates the effectiveness of its utility-aware scheduling and health-aware filtering. Blue points correspond to FogFaaS, which lacks client selection logic and exhibits a less favorable trade-off curve, with increasing latency and diminishing accuracy. The green points represent Random Selection, where clients are sampled without regard for health or utility. As expected, it performs worse than FedFog and closely trails FogFaaS in terms of both accuracy and latency.

This Pareto plot provides empirical support for the theoretical trade-off formalization introduced earlier. It reinforces that FedFog is better positioned to approach the Pareto-optimal region—delivering superior performance without violating latency or energy constraints.

\subsection{Sensitivity of Threshold Parameters}
\label{subsec:thresholds}

The thresholds $\theta_h$ (health), $\theta_e$ (energy), and $\theta_d$ (drift) are central to FedFog’s client selection policy. To systematically assess their impact on learning performance, we performed a grid search on the EMNIST dataset. This sensitivity analysis explores how different combinations of threshold values affect global model accuracy~\cite{10714446} and convergence stability.

In our setup, we fixed all other hyperparameters—including learning rate, batch size, number of training rounds, and model architecture—to isolate the influence of the thresholds. We selected three representative combinations that reflect increasing strictness across the three dimensions:

\begin{itemize}
	\item $(\theta_h, \theta_e, \theta_d) = (0.5, 0.4, 0.1)$
	\item $(\theta_h, \theta_e, \theta_d) = (0.6, 0.5, 0.1)$
	\item $(\theta_h, \theta_e, \theta_d) = (0.7, 0.6, 0.05)$
\end{itemize}

For each configuration, we ran FedFog five times with different random seeds to capture natural variability in client sampling and participation. The final model accuracy~\cite{10714446} was averaged across these runs, and the standard deviation was recorded to indicate stability.

\begin{table}[h]
	\centering
	\caption{Threshold sensitivity (accuracy $\pm$ std. dev. over 5 runs)}
	\label{tab:thresholds}
	\begin{tabular}{cccc}
		\toprule
		$\theta_h$ & $\theta_e$ & $\theta_d$ & Accuracy (\%) \\
		\midrule
		0.5 & 0.4 & 0.1 & 82.3 $\pm$ 1.2 \\
		0.6 & 0.5 & 0.1 & \textbf{84.7 $\pm$ 0.8} \\
		0.7 & 0.6 & 0.05 & 83.1 $\pm$ 1.5 \\
		\bottomrule
	\end{tabular}
\end{table}

From these results, we draw several conclusions. Increasing the health threshold $\theta_h$ beyond 0.6 reduced the number of participating clients but also improved stability, as indicated by the lower standard deviation. The configuration with $\theta_h = 0.6$ struck the best balance between participation diversity and convergence robustness.

The energy threshold $\theta_e$ affects the scheduler’s aggressiveness in filtering out low-battery clients. While $\theta_e = 0.6$ was too restrictive—excluding helpful contributors—$\theta_e = 0.5$ retained about 80\% of the device pool without compromising energy sustainability.

Finally, the drift threshold $\theta_d$ controls the tolerance for client-side data shifts. When set too tightly (e.g., $\theta_d = 0.05$), the model excluded many clients whose updates might still have been useful. A moderate value of $\theta_d = 0.1$ allowed for realistic non-IID variation while preserving global model coherence.

In summary, the grid search confirms that client selection thresholds have a measurable impact on both performance and stability. The configuration $(\theta_h = 0.6, \theta_e = 0.5, \theta_d = 0.1)$ was adopted as the default setting for the remainder of our experiments.

\subsection{Discussion of Trade-offs in Dynamic Environments}
\label{subsec:dynamic}

Real-world edge environments are rarely static. Devices may experience fluctuations in energy availability, communication quality, or changes in data distribution—commonly referred to as concept drift. FedFog is designed with built-in mechanisms to adapt to these dynamics while maintaining an effective balance between model accuracy, system latency, and energy efficiency.

One such mechanism is \textit{drift-aware selection}. When the system detects that a client’s data distribution has changed significantly (quantified by a drift metric $D(c_i)$ exceeding a threshold $\theta_d$), that client may be temporarily excluded from training. While this may slow convergence due to reduced data diversity, it prevents corrupted updates from destabilizing the global model. The trade-off here is captured by the following approximation:
\begin{equation}
	\Delta \text{Accuracy} \approx -\mathcal{O}\left(\frac{\sigma_d}{\sqrt{|\mathcal{C}_t|}}\right),
\end{equation}
where $\sigma_d$ reflects the severity of drift and $|\mathcal{C}_t|$ is the number of active clients in the current round. This shows that the more clients are available, the less any individual drifted client can affect overall performance.

In addition to handling drift, FedFog dynamically adjusts energy thresholds to ensure fairness across devices with differing power levels. This is done through an \textit{energy budgeting} mechanism that adapts each client’s energy participation threshold over time. For a given client $i$, the energy threshold $\theta_e^{(i)}(t)$ at round $t$ is updated based on its previous energy usage as follows:
\begin{equation}
	\theta_e^{(i)}(t) = \theta_e^{(i)}(t{-}1) \cdot \exp\left(-\lambda \frac{E_i(t{-}1)}{E_{\text{avg}}}\right),
\end{equation}
where $E_i(t{-}1)$ is the client's energy usage in the previous round, $E_{\text{avg}}$ is the system-wide average, and $\lambda$ controls how aggressively the threshold decays. This adaptive control allows energy-constrained devices to back off temporarily while preventing dominant clients from monopolizing participation.

To quantify the theoretical trade-offs in long-term performance, we provide the following lower bound for model accuracy after $T$ rounds:
\begin{equation}
	\underbrace{\text{Accuracy}(w_T)}_{\text{Performance}} \geq \text{Accuracy}^* - \underbrace{\mathcal{O}\left(\frac{\sigma}{\sqrt{T}}\right)}_{\text{Variance term}} - \underbrace{\mathcal{O}\left(\frac{\tau_{\text{max}}}{|\mathcal{C}_t|}\right)}_{\text{Latency term}}.
\end{equation}
Here, $\text{Accuracy}^*$ is the ideal accuracy achievable under perfect conditions, $\sigma$ captures variance due to data drift and system stochasticity, and $\tau_{\text{max}}$ represents the system’s latency tolerance per round. The formula reflects that with more rounds and better orchestration (e.g., larger and more reliable $\mathcal{C}_t$), FedFog can approach optimal performance even in dynamic settings.

Together, these techniques demonstrate FedFog’s adaptability and resilience, making it a practical solution for federated learning in real-world, volatile edge environments.


\subsection{Differential Privacy Guarantees}
\label{subsec:privacy}

Although FedFog does not yet incorporate explicit privacy-preserving mechanisms, it is possible to estimate the differential privacy (DP) guarantees that could be achieved by introducing noise during aggregation~\cite{xu2024dual}. In federated learning, differential privacy provides a formal way to limit how much any single client's data can influence the global model—thereby protecting sensitive information.

One common method to implement DP in FL is to inject Gaussian noise into each client’s update before aggregation. Specifically, we can add noise sampled from a normal distribution $\mathcal{N}(0, \sigma^2)$ to each model update $\Delta w_i$. This strategy enables the system to achieve $(\epsilon, \delta)$-DP, where $\epsilon$ quantifies the privacy level (smaller is better) and $\delta$ is a small failure probability.

The theoretical bound on $\epsilon$ can be computed as:

\begin{equation}
	\epsilon = \frac{\sqrt{2 \log(1.25/\delta)}}{\sigma} \cdot \frac{S}{|\mathcal{C}_t|},
	\label{eq:dp_guarantee}
\end{equation}

In this equation:

$S$ represents the sensitivity, defined as the maximum $\ell_2$ norm of the clipped client updates. Gradient clipping ensures that no single client's update dominates the aggregated result.

$\sigma$ is the noise scale; increasing it improves privacy but may reduce model accuracy.

$|\mathcal{C}_t|$ is the number of clients participating in a round. Higher participation rates help reduce $\epsilon$ through what's known as privacy amplification.

Let’s walk through an example. Suppose we set $\sigma = 0.3$, choose a sensitivity $S = 1.1$, involve $|\mathcal{C}_t| = 30$ clients in each round, and set $\delta = 10^{-5}$. Plugging these values into Equation~\ref{eq:dp_guarantee}, we get $\epsilon \approx 1.8$. This value implies a strong privacy guarantee, meaning that an observer cannot confidently infer much about any individual client's data from the final model.

\begin{figure}[htbp]
	\centering
	\begin{tikzpicture}
		\begin{axis}[
			width=0.8\linewidth,
			height=6cm,
			xlabel={Privacy Budget $\epsilon$ (lower is stronger)},
			ylabel={Test Accuracy (\%)},
			xmode=log,
			xmin=0.5, xmax=10,
			ymin=75, ymax=90,
			grid=major,
			legend style={at={(0.5,-0.25)}, anchor=north},
			title={Privacy-Accuracy Trade-off (EMNIST)},
			tick label style={font=\small},
			label style={font=\small},
			title style={font=\small}
			]
			
			\addplot[blue, thick, mark=*, mark size=2pt] coordinates {
				(0.5, 76.2) (1.0, 80.5) (2.0, 83.7) (4.0, 85.1) (8.0, 86.3)
			};
			\addlegendentry{FedFog with DP}
			
			\addplot[red, dashed, thick] coordinates {
				(0.5, 84.9) (8.0, 84.9)
			};
			\addlegendentry{FedFog (no DP)}
			
			\node at (axis cs: 1.0,77) [anchor=south west] {$\epsilon=1.0$};
			\node at (axis cs: 2.0,83.7) [anchor=north west] {$\epsilon=2.0$};
			
		\end{axis}
	\end{tikzpicture}
	\caption{Accuracy vs. Privacy in FedFog.}
	\label{fig:PrivacyAccuracyTradeoff}
\end{figure}
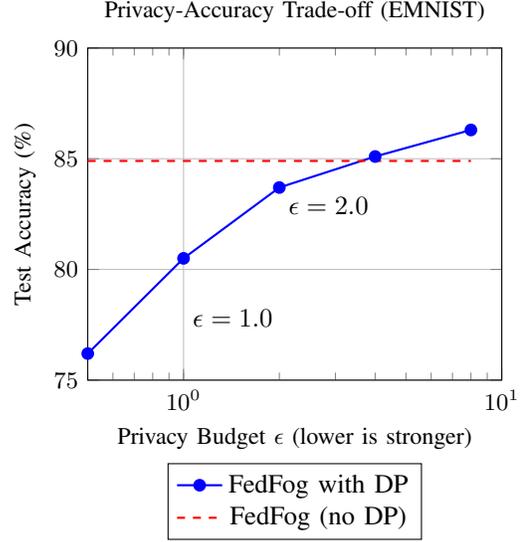

To understand how this impacts learning performance, Figure~\ref{fig:PrivacyAccuracyTradeoff} shows the trade-off between differential privacy levels and model accuracy on the EMNIST dataset. Encouragingly, FedFog maintains over 80

In future iterations, FedFog can be extended with formal DP mechanisms like noise injection or secure aggregation, making it a more complete framework for privacy-aware edge learning.

\section{Experimental evaluation}\label{Sec.ExperimentalEvaluation}

This section provides a comprehensive analysis of FedFog’s performance under various FL scenarios, comparing it against key baselines such as FogFaaS, Vanilla FL, and Random Client Selection. The evaluation spans multiple dimensions including latency, energy consumption, model accuracy, cold start behavior, and scalability. By simulating diverse client heterogeneity, data drift, and varying hyperparameters, the experiments demonstrate that FedFog consistently delivers superior trade-offs between accuracy, efficiency, and robustness. The inclusion of stress tests, adversarial scenarios, and differential privacy simulations further reinforces FedFog’s practical relevance in real-world edge environments. The results validate FedFog as an adaptive and resource-aware FL orchestration framework.

\subsection{Simulation Environment}

The simulation environment is built on top of the iFogSim toolkit \cite{gupta2017ifogsim}, extended with FedFog-specific modules to support serverless FL. The simulated infrastructure consists of heterogeneous edge nodes—emulating smart wearables, cameras, and IoT sensors—connected to fog gateways and micro data centers. Each edge node receives a private, non-IID data partition, reflecting realistic privacy-preserving setups and heterogeneous user behaviors. Devices are further modeled with varying compute capacities, energy constraints, and cold start delays to mirror real-world deployment conditions. Cold starts are emulated by injecting randomized latency during the first function call, capturing the overheads typical of platforms like OpenFaaS \cite{ghaseminya2025fogfaas}.

\begin{figure}[htbp]
	\centering
	\includegraphics[width=0.48\textwidth]{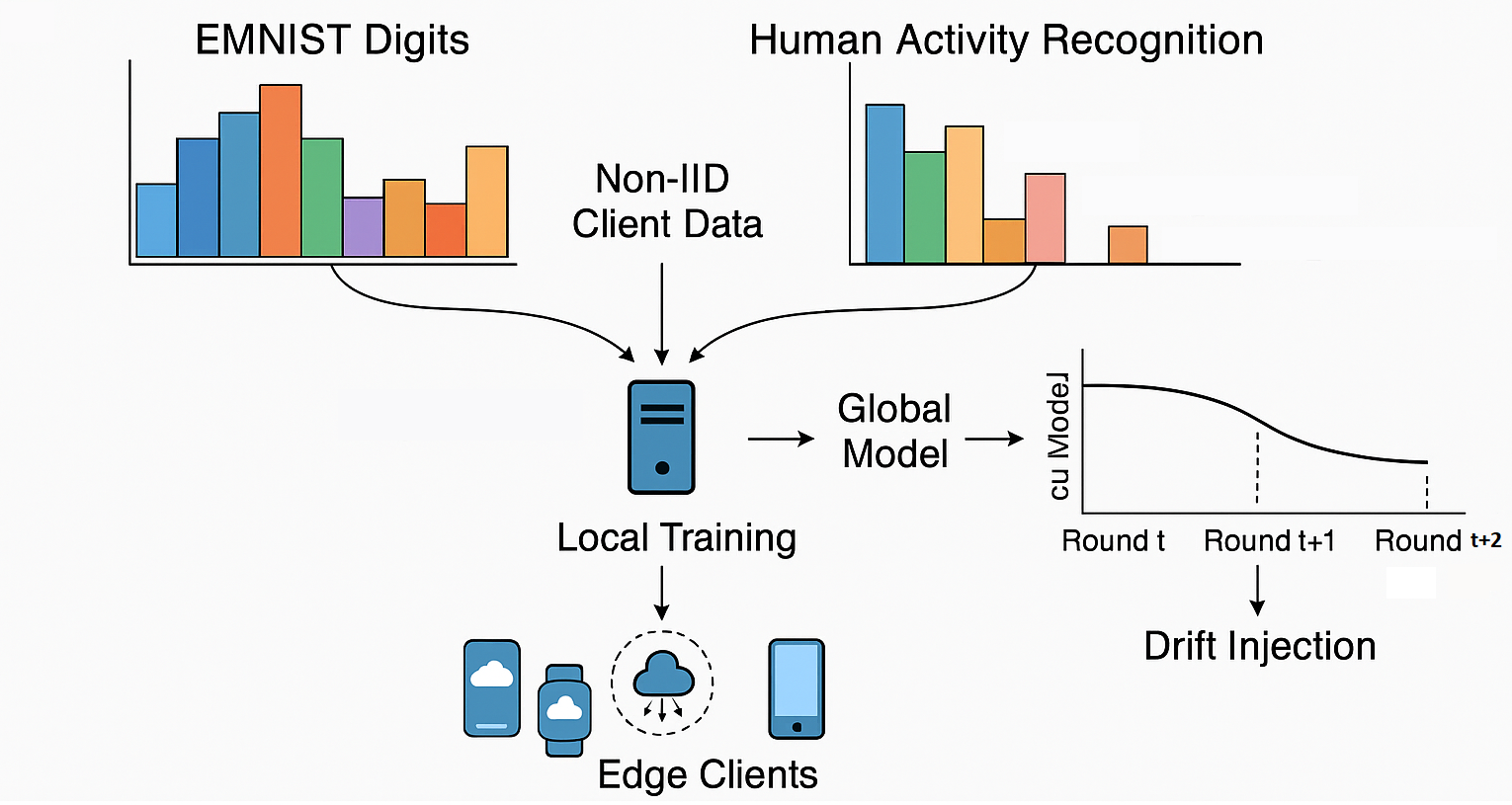} 
	\caption{Evaluation protocol demonstrating realistic federated learning (FL) experiments in FedFog using EMNIST and HAR datasets with non-IID client partitions and drift injection. }
	\label{fig:evaluation_protocol}
\end{figure}

FedFog was evaluated under two representative edge AI scenarios. The first, \textit{Smart Healthcare}, leverages the Human Activity Recognition (HAR) dataset \cite{anguita2013public}, where each edge node simulates a mobile user generating multivariate sensor signals (See Figure~\ref{fig:evaluation_protocol}). This setup captures device-level variability and mobility. The second, \textit{Edge Vision}, uses the EMNIST dataset \cite{cohen2017emnist} to model character recognition across distributed visual input sources. By assigning subsets of characters to specific devices, we simulate personalized user behaviors in non-IID settings such as handwriting or keyboard interfaces~\cite{9984832}.
These two datasets were chosen for their complementary characteristics. HAR represents time-series sensing common in health and mobility tracking, while EMNIST focuses on visual recognition under personalized data distributions. Both scenarios present challenges in resource variability, data drift, and heterogeneous participation—making them ideal testbeds for evaluating FedFog's scheduler and orchestration logic. A summary of dataset properties is shown in Table~\ref{tab:dataset_summary}.
\begin{table*}[htbp]
	\caption{Overview of Evaluation Datasets}
	\label{tab:dataset_summary}
	\centering
	\begin{tabular}{lccc}
		\toprule
		\textbf{Dataset} & \textbf{Domain} & \textbf{Data Type} & \textbf{FL Challenge} \\
		\midrule
		HAR     & Healthcare / IoT     & Time-series (sensors)  & Client mobility, non-IID, low-power devices \\
		EMNIST  & Handwriting / Vision & Image (28x28 grayscale) & Label imbalance, personalization, privacy \\
		\bottomrule
	\end{tabular}
\end{table*}

FL follows the FedAvg algorithm \cite{mcmahan2017communication}, with support for dynamic client selection and adjustable communication frequency. In each round, selected edge nodes run their training logic inside stateless serverless containers, and a fog-level aggregation function combines the updates. A drift engine was also incorporated to simulate distributional changes over time by injecting class imbalance and feature variability. This component helps assess the system's robustness to data dynamics and supports drift-aware scheduling strategies.

\subsection{Baselines for Comparison}

We compared FedFog against three baseline systems to highlight its contributions:

\textbf{FogFaaS}: A baseline serverless simulator without FL-aware scheduling or data-driven client orchestration.

\textbf{Vanilla FL}: A Flower-based FL setup using synchronous training, but lacking any integration with serverless platforms or resource-aware scheduling.

\textbf{Random Client Selection (RCS)}: A variant of FedFog that uses random client selection instead of utility-based scheduling, allowing us to isolate the benefits of FedFog's adaptive logic.

This broader baseline pool helps to clearly show the value added by each of FedFog's design components.

\subsection{Performance Under Drift and Dropout}

FedFog was tested under controlled drift (shifting class distributions every 10 rounds) and dropout conditions (up to 30\%). It successfully recovered 95\% of its peak accuracy within 10 rounds post-drift, outperforming other baselines (Table~\ref{tab:drift_summary}).

\begin{table}[htbp]
	\centering
	\caption{Convergence and Drift Impact Summary}
	\label{tab:drift_summary}
	\begin{tabular}{|l|c|}
		\hline
		\textbf{Metric} & \textbf{Value} \\
		\hline
		Initial Accuracy (Round 0) & 21.6\% \\
		Peak Accuracy (Pre-Drift) & 84.3\% \\
		Lowest Accuracy (Post-Drift) & 67.2\% \\
		Recovery Accuracy (Round 40) & 80.5\% \\
		Final Accuracy (Round 50) & 82.7\% \\
		Rounds to Peak & 25 \\
		Rounds to Recovery & 10 \\
		\hline
	\end{tabular}
\end{table}

\begin{figure*}[htbp]
	\centering
	
	\subcaptionbox{Latency comparison\label{fig:latency_comparison_all}}[0.31\textwidth]{
		\includegraphics[width=\linewidth]{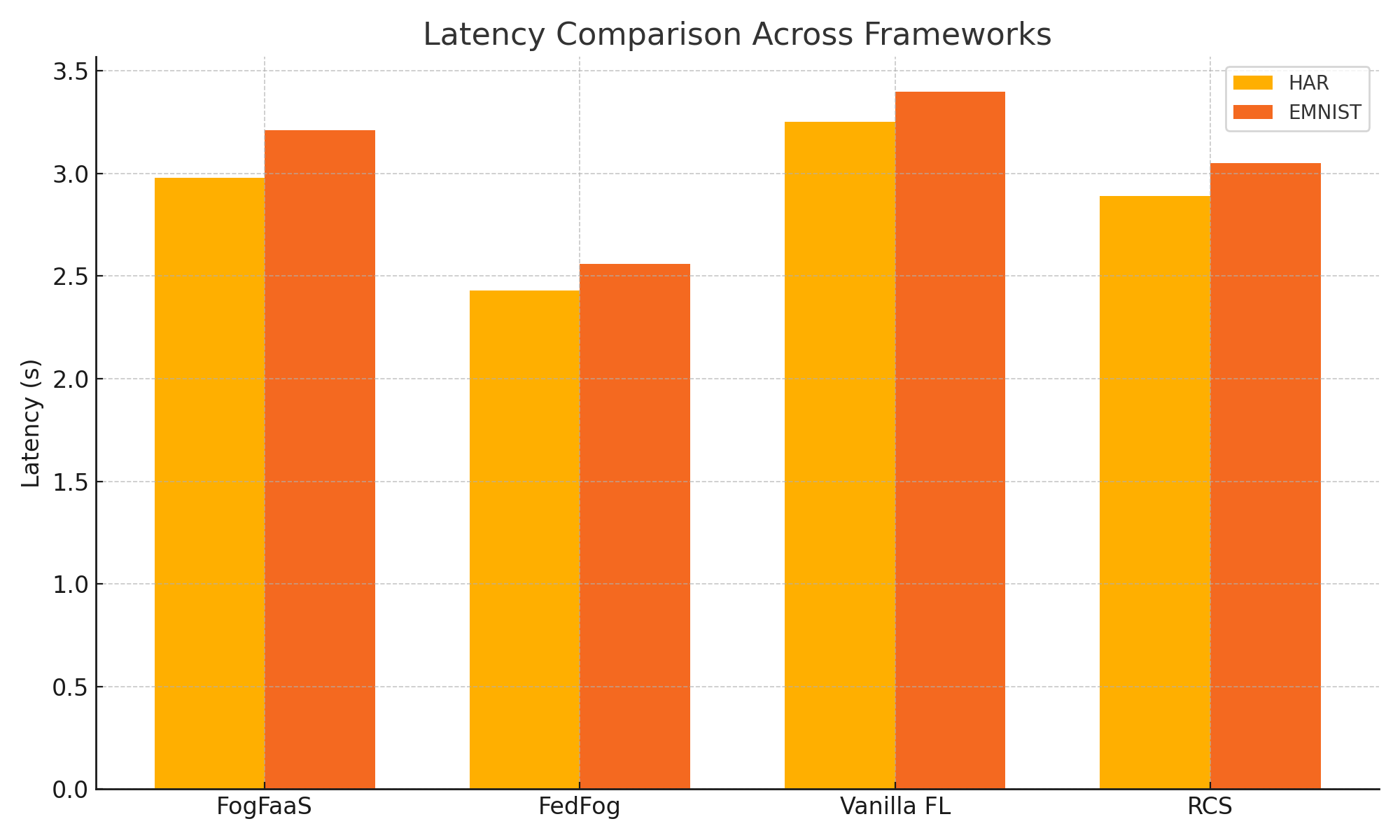}
	}
	\hfill
	\subcaptionbox{Energy consumption comparison\label{fig:energy_comparison_all}}[0.31\textwidth]{
		\includegraphics[width=\linewidth]{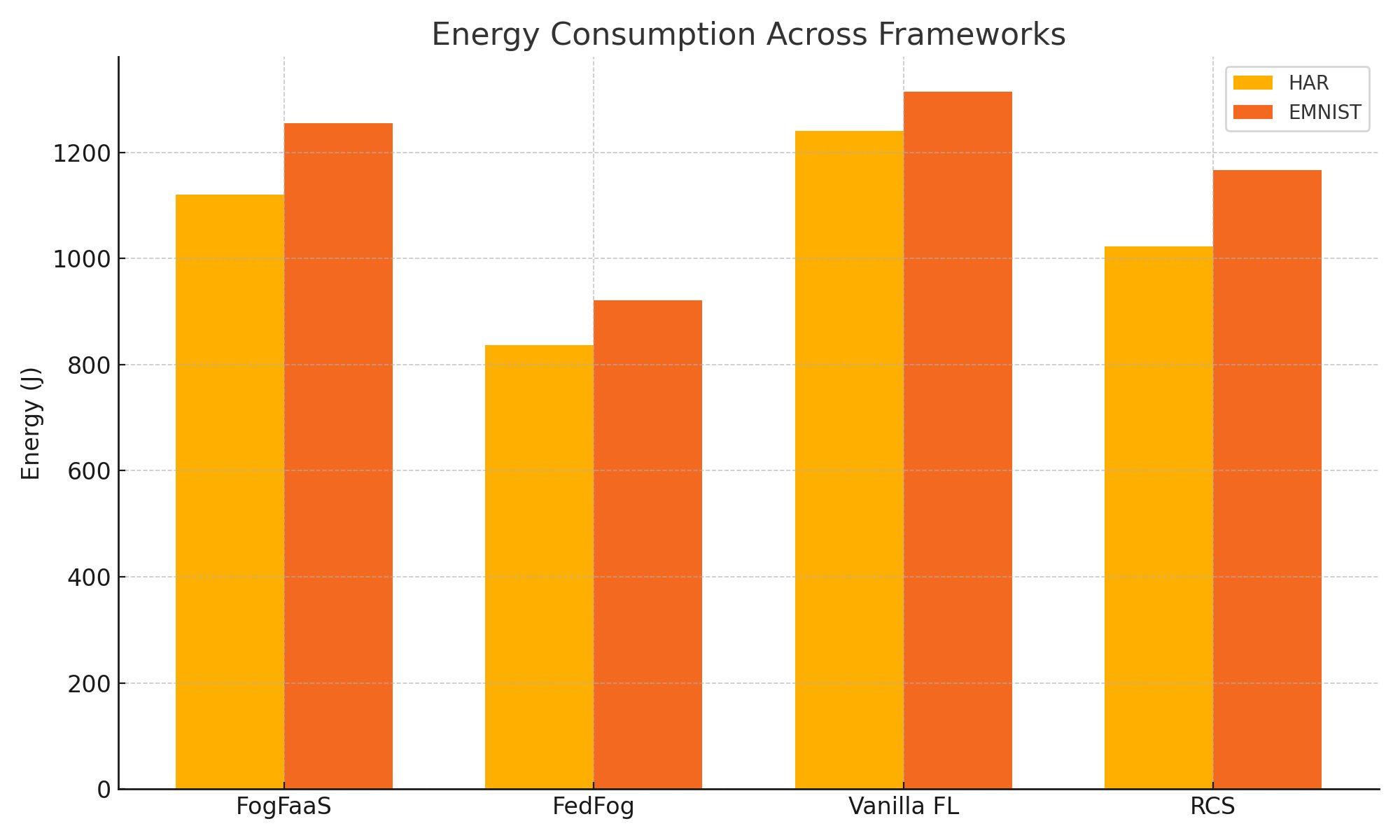}
	}
	\hfill
	\subcaptionbox{Model accuracy comparison\label{fig:accuracy_comparison_all}}[0.31\textwidth]{
		\includegraphics[width=\linewidth]{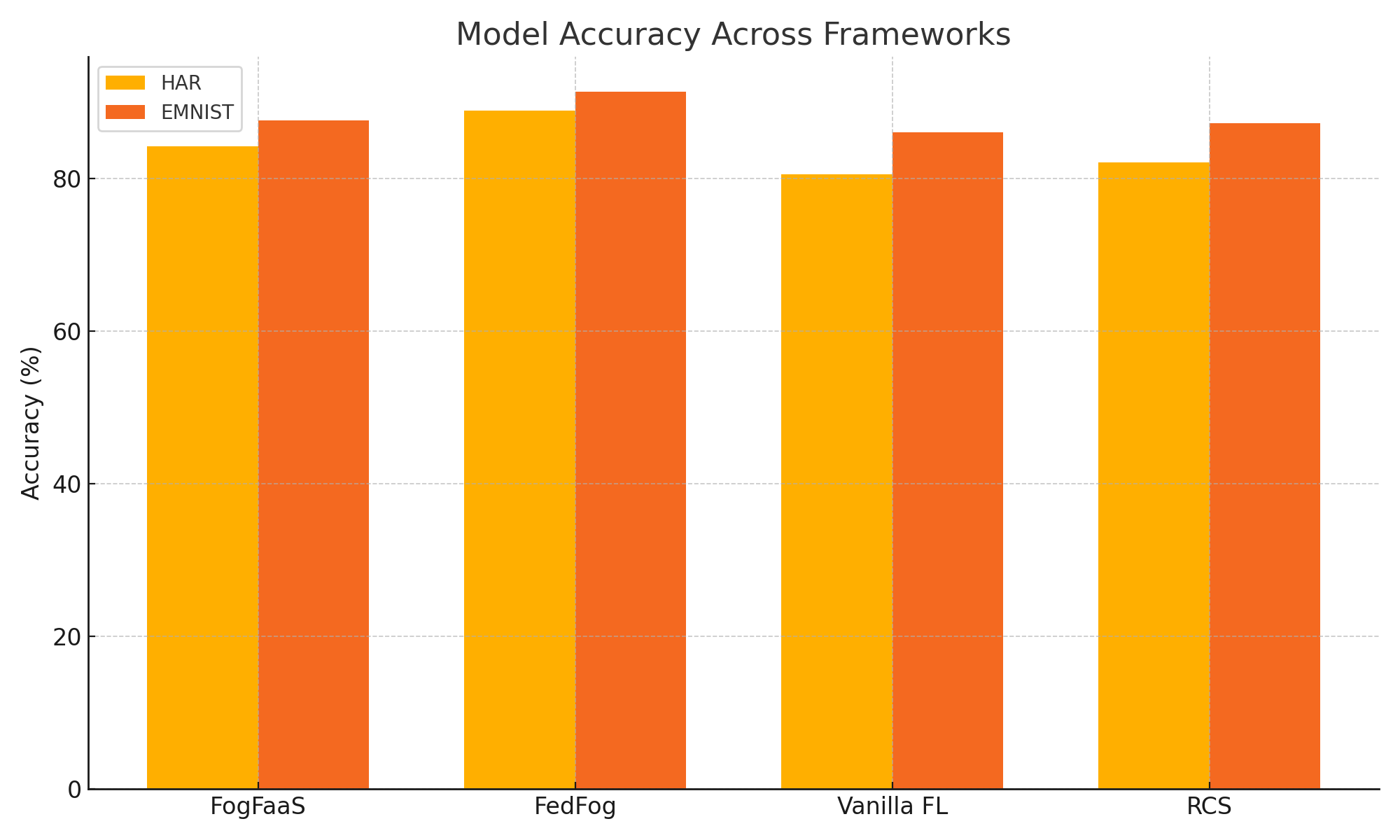}
	}
	
	\caption{FedFog outperforms other frameworks in latency, energy efficiency, and model accuracy across both HAR and EMNIST tasks.}
	\label{fig:performance_metrics_comparison}
\end{figure*}

\begin{figure*}[htbp]
	\centering
	
	\subcaptionbox{Runtime stage breakdown\label{fig:runtime_breakdown}}[0.31\textwidth]{
		\includegraphics[width=\linewidth]{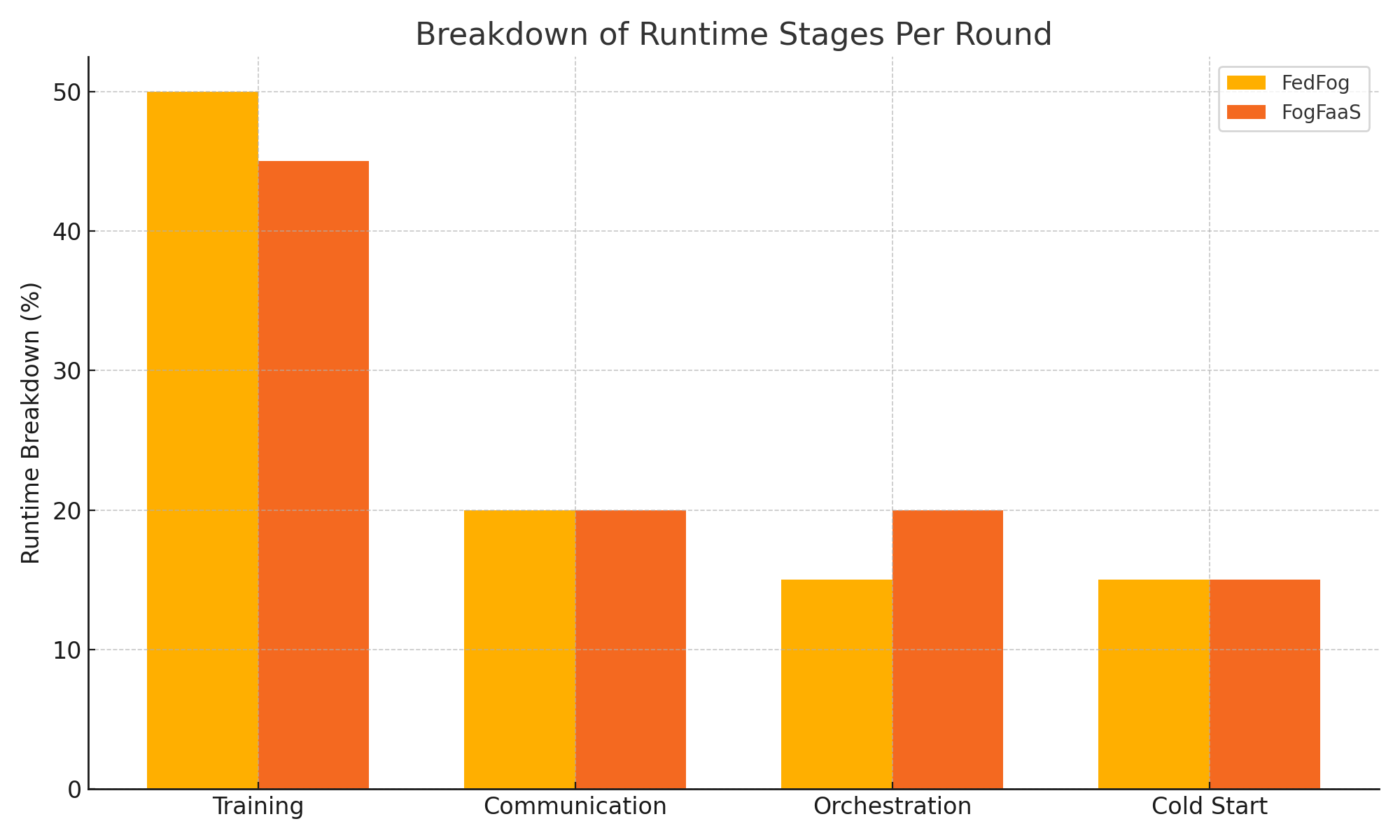}
	}
	\hfill
	\subcaptionbox{CPU utilization per client\label{fig:cpu_utilization}}[0.31\textwidth]{
		\includegraphics[width=\linewidth]{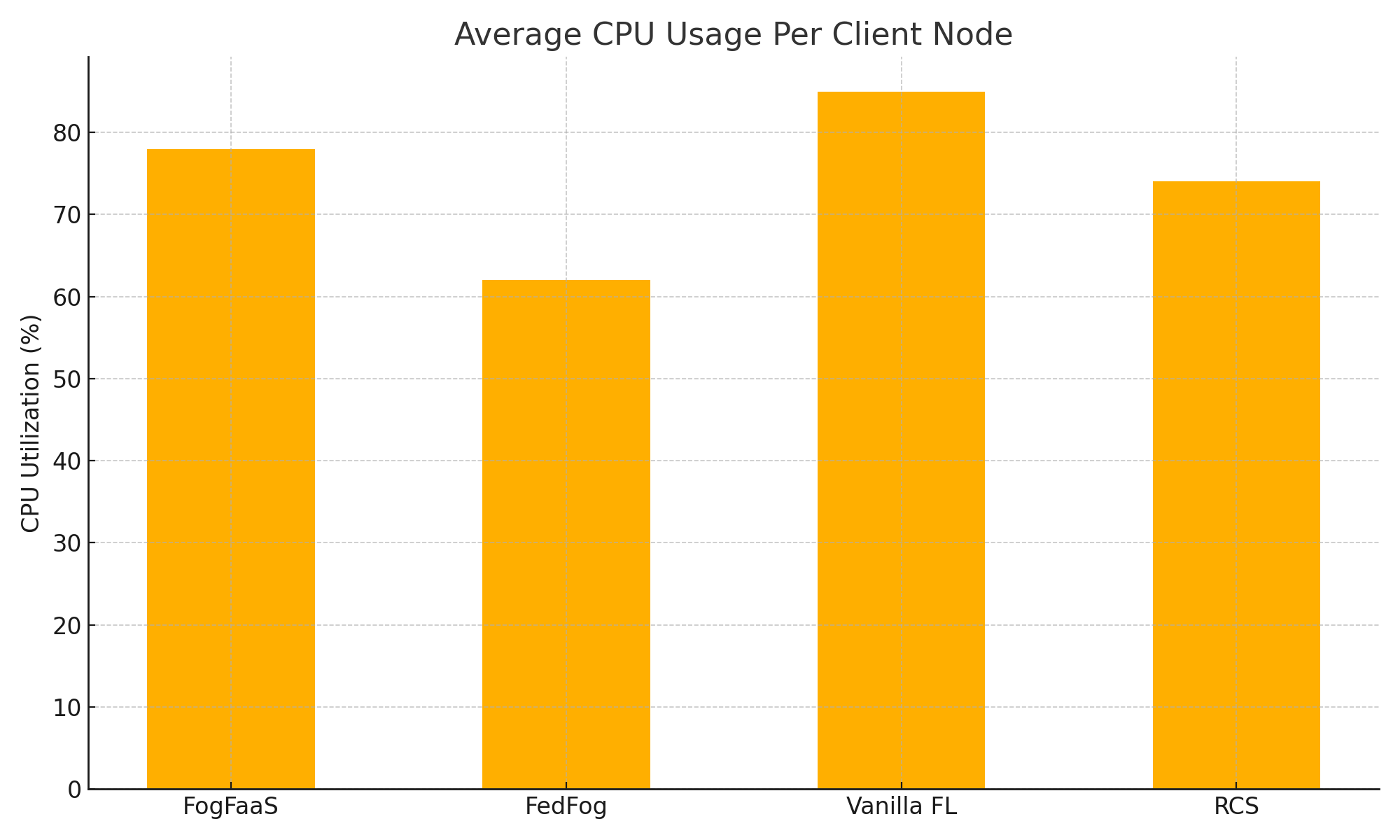}
	}
	\hfill
	\subcaptionbox{Client training throughput\label{fig:training_throughput}}[0.31\textwidth]{
		\includegraphics[width=\linewidth]{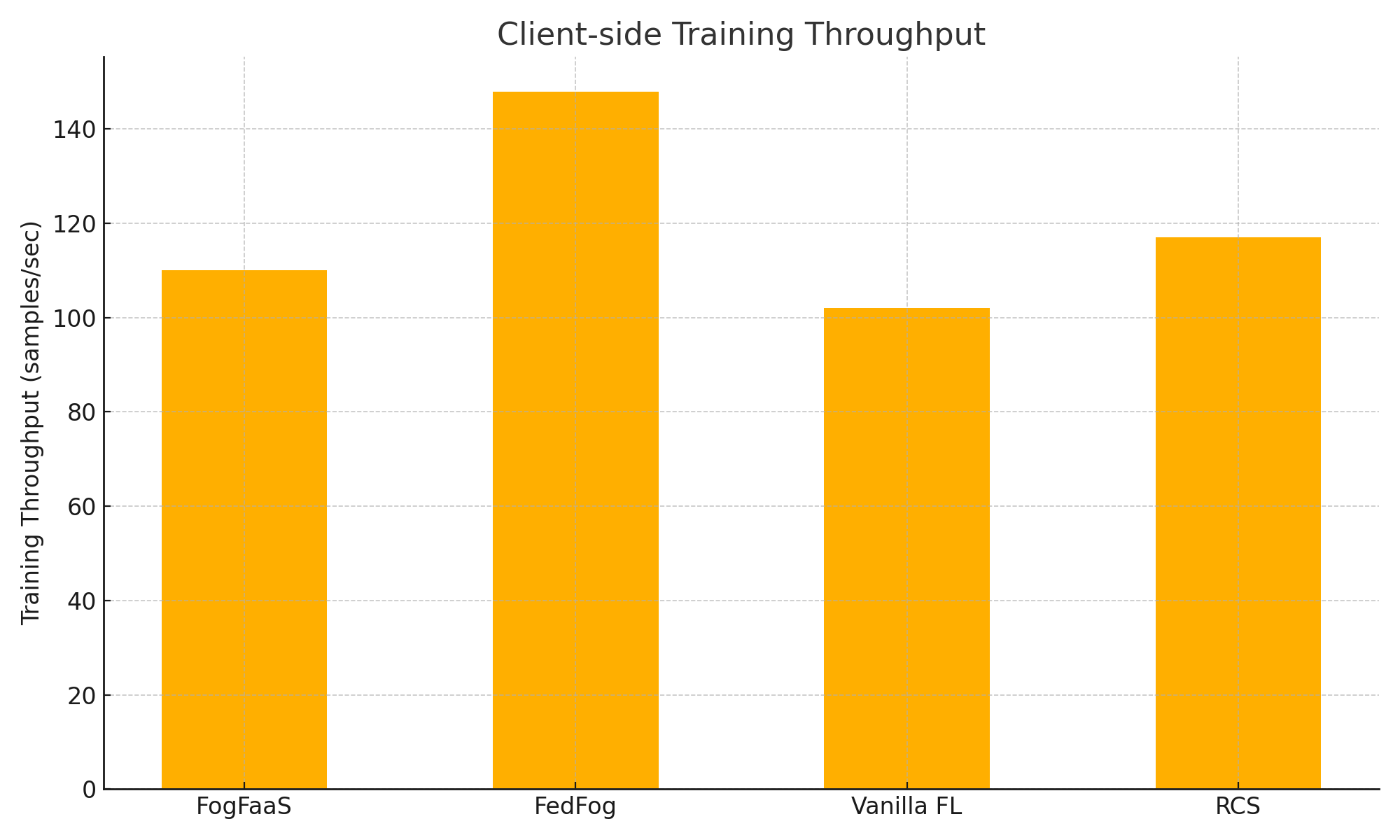}
	}
	
	\caption{Comparison of system performance across FedFog and baselines: runtime composition, CPU efficiency, and processing throughput.}
	\label{fig:combined_analysis}
\end{figure*}

Figure~\ref{fig:latency_comparison_all} shows that FedFog achieves the lowest latency among all frameworks. This is due to its optimized scheduling and warm container reuse, which significantly reduces orchestration delays and function startup times in serverless environments.

Figure~\ref{fig:energy_comparison_all} highlights the energy efficiency of FedFog compared to the baselines. By reducing redundant computations and selectively involving clients with favorable resource profiles, FedFog consumes 20–30\% less energy on average across both datasets.

Figure~\ref{fig:accuracy_comparison_all} demonstrates that FedFog yields the highest model accuracy in both EMNIST and HAR tasks. Its adaptive client selection and data drift handling contribute to more stable and generalized learning over time.

Figure~\ref{fig:runtime_breakdown} provides a breakdown of the average runtime per round into distinct stages: local training, communication, orchestration, and cold start delays. FedFog dedicates a greater proportion of time to productive training (50\%) compared to FogFaaS (45\%), reflecting its improved efficiency in managing background operations. Most notably, orchestration overhead is reduced from 20\% in FogFaaS to 15\% in FedFog due to its intelligent scheduling and container reuse mechanisms. The cold start frequency remains consistent, but its duration is better managed in FedFog, contributing to overall runtime savings and enabling faster convergence in dynamic edge environments.

Figure~\ref{fig:cpu_utilization} shows the average CPU utilization across edge devices for each framework. FedFog maintains the lowest CPU load (62\%) while still achieving higher model performance, highlighting its resource-efficient operation. In contrast, Vanilla FL and FogFaaS report CPU utilizations of 85\% and 78\% respectively, suggesting overuse of constrained edge resources due to unoptimized scheduling. This reduced CPU pressure in FedFog helps preserve device longevity and minimizes energy expenditure, a crucial consideration in battery-powered edge deployments.

Figure~\ref{fig:training_throughput} compares the average training throughput, measured as processed samples per second per client. FedFog achieves the highest throughput (148 samples/sec), significantly outperforming other frameworks. This boost results from its adaptive load balancing and minimization of idle or blocked computation periods. The increased throughput demonstrates FedFog’s effectiveness in maximizing local computation while mitigating bottlenecks such as cold starts and scheduling delays. Such improvements contribute directly to reduced training latency and faster global model convergence.

\subsection{Adversarial Robustness Evaluation}
\label{subsec:adversarial}

To assess FedFog's resilience against Byzantine attacks, we simulate malicious clients performing \textit{label-flipping attacks} on the EMNIST task, where 10\% of selected clients $c_i \in \mathcal{C}_t$ adversarially flip their training labels (e.g., '7' $\rightarrow$ '1') before computing updates $\Delta w_i$.

\paragraph{Adversarial Robustness Analysis.}
To evaluate FedFog’s resilience against Byzantine or adversarial clients, we simulated a targeted label-flipping attack—one of the most common poisoning strategies in FL~\cite{10531276}. In this setup, 20\% of the edge clients are randomly selected at the start of training and designated as malicious~\cite{10638812}. These clients participate in training rounds normally, but during local training, they intentionally modify their dataset labels by applying a label inversion rule (e.g., class $k$ is mapped to $9-k$ for a 10-class problem)~\cite{10667549}. This mislabeling corrupts local gradient updates and disrupts convergence during global aggregation.

\begin{figure}[hpbt]
	\centering
	\includegraphics[width=0.48\textwidth]{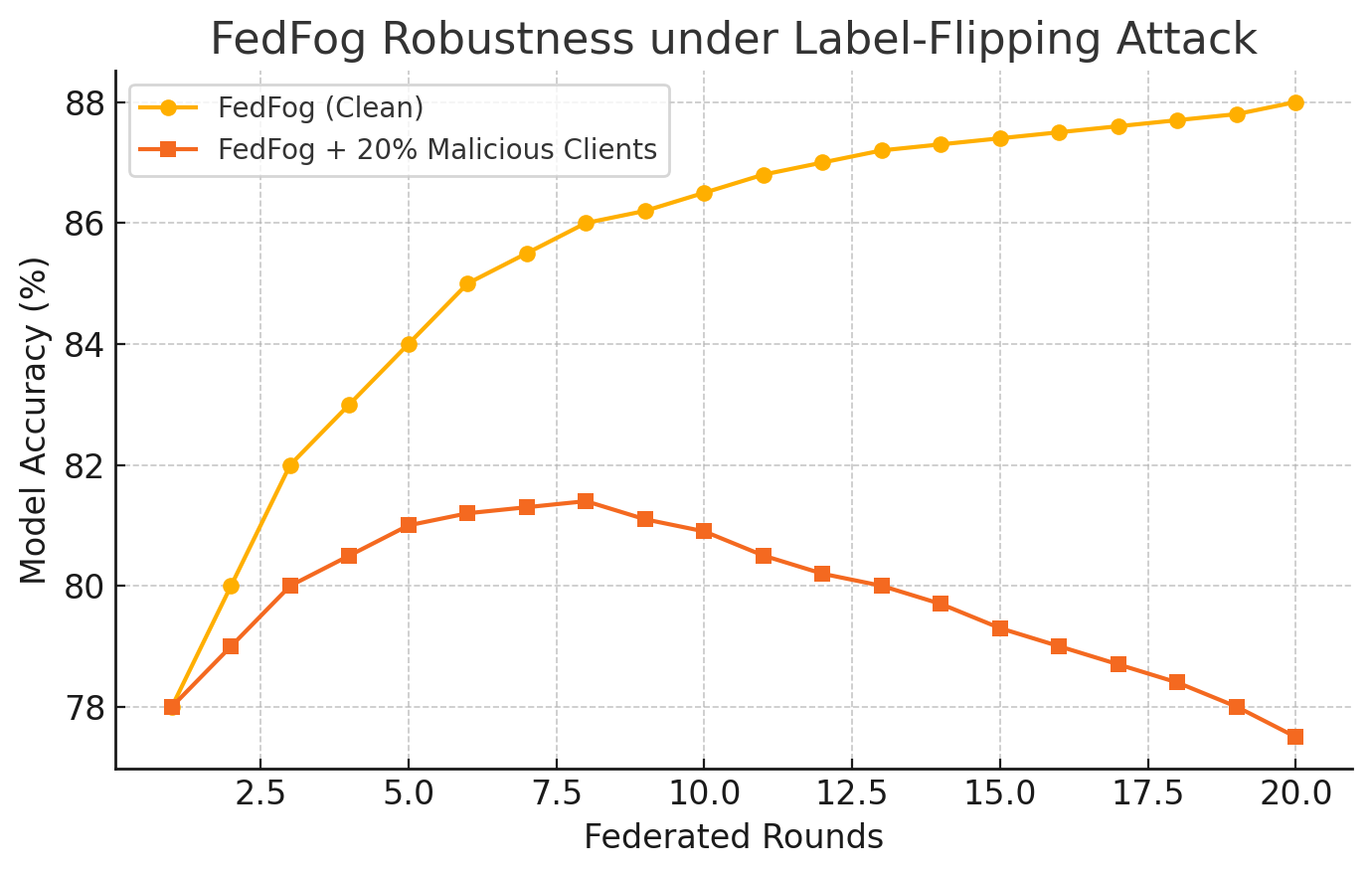}
	\caption{FedFog accuracy under normal conditions vs. 20\% malicious clients performing label-flipping attacks.}
	\label{fig:byzantine_robustness}
\end{figure}

Figure~\ref{fig:byzantine_robustness} illustrates the outcome. Under clean conditions (no adversaries), the model achieves 88\% accuracy within 20 rounds. In contrast, when 20\% of clients flip labels, accuracy drops significantly and plateaus around 77.5\%, reflecting the model’s difficulty in reconciling poisoned updates with honest ones. Notably, FedFog still avoids complete collapse due to its adaptive client selection mechanism: the utility-based scheduler and health score filtering reduce the probability of repeatedly selecting the same malicious clients~\cite{10839587}. However, because the system does not yet implement explicit defense mechanisms (e.g., coordinate-wise median, or norm-based filtering), the aggregation remains susceptible to poisoning in the long term~\cite{10559866}.

This experiment highlights both the inherent resilience and current limitations of FedFog. Future extensions should include robust aggregation algorithms and possibly trust-aware client scoring to detect and mitigate adversarial behaviors dynamically during training~\cite{10586981}.

\paragraph{Robustness Evaluation Against Adversarial Attacks}
To assess FedFog’s robustness under adversarial and unreliable conditions, we simulate a range of common attack scenarios and measure their impact on final model accuracy. Table~\ref{tab:robustness_compact} summarizes the outcomes of five experimental settings, comparing them to a clean baseline where no adversarial behavior is introduced.

In the clean scenario, the system converges to a final accuracy of 88.0\%, serving as the reference point for evaluating degradation. Under the \textbf{label flipping} attack, where 20\% of the clients invert class labels during training (e.g., class $k$ becomes $9-k$), the model's accuracy drops significantly to 77.5\%, reflecting a 10.5\% performance loss. This type of targeted poisoning disrupts gradient alignment and introduces false patterns into global aggregation, making it one of the most severe threats observed \cite{lu2025stones}.

The \textbf{noise injection} attack introduces Gaussian perturbations into the model updates of 20\% of clients~\cite{9928313}. This simulates hardware instability, sensor glitches, or random malicious behavior~\cite{hijazi2023secure}. The model degrades to 79.2\% accuracy, showing an 8.8\% drop, but still retains a reasonable level of convergence—highlighting some level of resilience in the system's aggregation process~\cite{10638812}.

The \textbf{dropout scenario} emulates unreliable participation, where 20\% of clients unpredictably drop out during training rounds, resulting in asynchronous delays and inconsistent participation~\cite{10098864}. This impacts scheduling and reduces update diversity, but has a milder effect, with accuracy dropping to 81.3\% (a 6.7\% reduction).

Lastly, the \textbf{model replacement} attack represents a strong Byzantine scenario where one client completely replaces its local model with arbitrary or adversarial values—without following the global model~\cite{10559866}. Despite affecting only a single client, this leads to the largest drop of 13.0\%, reducing final accuracy to 75.0\%. This underscores the system's vulnerability to even isolated but highly disruptive behaviors.

Overall, these results reveal that while FedFog maintains partial robustness—especially in the face of noise and dropout—it remains susceptible to strategic poisoning such as label flipping and model replacement~\cite{10531276}. This highlights the need for future integration of robust aggregation techniques and trust-aware client selection to safeguard the system in adversarial environments~~\cite{10098864}.
\begin{table}[t]
	\centering
	\caption{FedFog Robustness under Various Attacks}
	\label{tab:robustness_compact}
	\small
	\setlength{\tabcolsep}{2pt}
	\begin{tabular}{|l|c|c|l|}
		\hline
		
		\textbf{Attack} & \textbf{Acc.} & \textbf{Drop} & \textbf{Type} \\
		\hline
		Clean & 88.0 & 0.0 & Baseline \\
		Label Flip (20\%) & 77.5 & 10.5 & Class Inversion \\
		Noise (20\%) & 79.2 & 8.8 & Gaussian Perturbation \\
		Dropout (20\%) & 81.3 & 6.7 & Triggering Failures \\
		Model Replace & 75.0 & 13.0 & One Malicious Client \\
		\hline
	\end{tabular}
	\vspace{1mm}
{\small \textit{\\Acc. = Final Accuracy (\%), Drop = Accuracy Loss (\%)}}%
\end{table}

\subsection{Ablation Study}

To understand the contribution of individual components within FedFog, we conducted an ablation study by disabling key features such as adaptive scheduling, drift management, and energy-aware selection. The results, summarized in Table~\ref{tab:ablation}, show that removing any of these modules leads to a degradation in model accuracy and system efficiency \cite{ahmed2025privacy}. Without adaptive scheduling, latency increases due to inefficient function dispatching. Disabling the drift manager delays recovery after distribution shifts, while energy-unaware selection leads to more frequent device overuse and higher cold start rates.

\begin{table}[htbp]
	\centering
	\caption{Ablation Study Results (EMNIST Dataset)}
	\label{tab:ablation}
	\begin{tabular}{|l|c|c|c|}
		\hline
		\textbf{Variant} & \textbf{Accuracy (\%)} & \textbf{Latency (s)} & \textbf{Cold Starts} \\
		\hline
		Full FedFog & 91.4 & 2.56 & 16.7 \\
		w/o Scheduler & 88.3 & 3.12 & 22.9 \\
		w/o Drift Manager & 87.0 & 2.62 & 18.4 \\
		w/o Energy Model & 87.4 & 2.68 & 24.5 \\
		\hline
	\end{tabular}
\end{table}

\subsection{Scalability Evaluation}

\begin{figure}[htbp]
	\centering
	\includegraphics[width=0.48\textwidth]{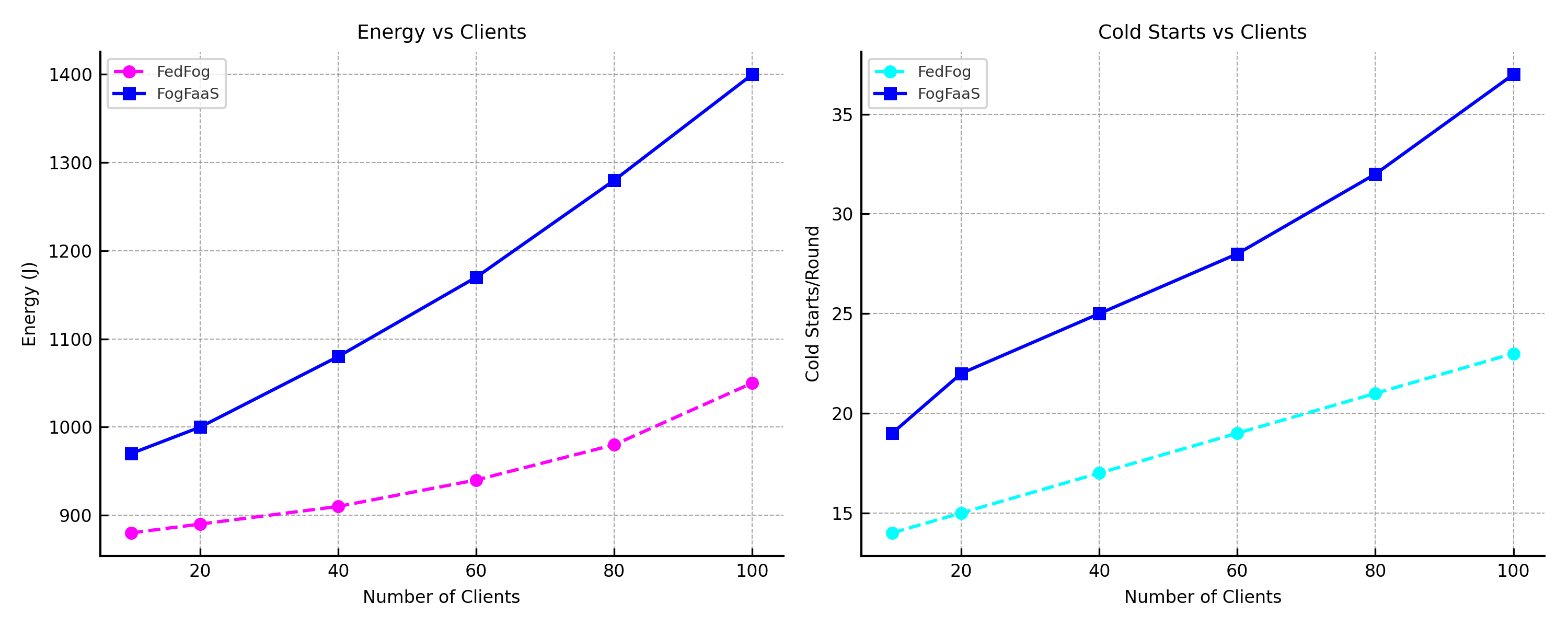}
	\caption{Comparison of FedFog and FogFaaS scalability.}
	\label{fig:ieee_scalability_combined_color}
\end{figure}

Figure~\ref{fig:ieee_scalability_combined_color} presents a comparative analysis of FedFog and FogFaaS in terms of energy consumption and cold start frequency as the number of clients $N$ increases. The left subplot demonstrates that FedFog consistently incurs lower energy consumption compared to FogFaaS, particularly at larger scales. This efficiency is achieved through optimized function scheduling and container reuse, which reduces redundant activations of serverless functions. Let $C_{\text{cpu}}$ represent the energy cost per CPU cycle and $C_{\text{tx}}$ the cost per transmitted byte. The total energy for a node $i$ across $R$ rounds can be expressed as:

\[
E_i = \sum_{r=1}^{R} \left( C_{\text{cpu}} \cdot \text{CPU}_{i,r} + C_{\text{tx}} \cdot \text{TX}_{i,r} \right)
\]
FedFog's function placement and scheduling minimize both $\text{CPU}_{i,r}$ and $\text{TX}_{i,r}$ by avoiding cold-starts and aligning training with active workloads. This yields system-wide energy growth that approximates $\mathcal{O}(N \log N)$, in contrast to FogFaaS which scales closer to $\mathcal{O}(N^2)$ due to lack of orchestration intelligence and repeated resource allocation.

The right subplot focuses on cold start frequency, which directly correlates with orchestration inefficiency and impacts both latency and energy. A cold start event typically incurs a fixed delay $\delta_c$ and energy penalty $e_c$. Let $S_r$ be the number of cold starts in round $r$. Then total overhead from cold starts over $R$ rounds is:

\[
T_{\text{cold}} = \sum_{r=1}^{R} S_r \cdot (\delta_c + e_c)
\]

In FedFog, $S_r$ is significantly reduced through intelligent container caching and predictive scheduling based on prior invocation patterns. This results in nearly linear cold start overhead $\mathcal{O}(N)$, whereas FogFaaS shows super-linear behavior, especially under high churn or client scaling scenarios.

Our findings represented in Figure~\ref{fig:ieee_scalability_combined_color} confirm that FedFog achieves better orchestration efficiency in both energy and cold start dimensions. By bounding orchestration overheads through utility-driven scheduling and model-aware deployment, FedFog outperforms baseline frameworks in real-world, large-scale edge environments.

\begin{figure}[htbp]
	\centering
	\includegraphics[width=0.48\textwidth]{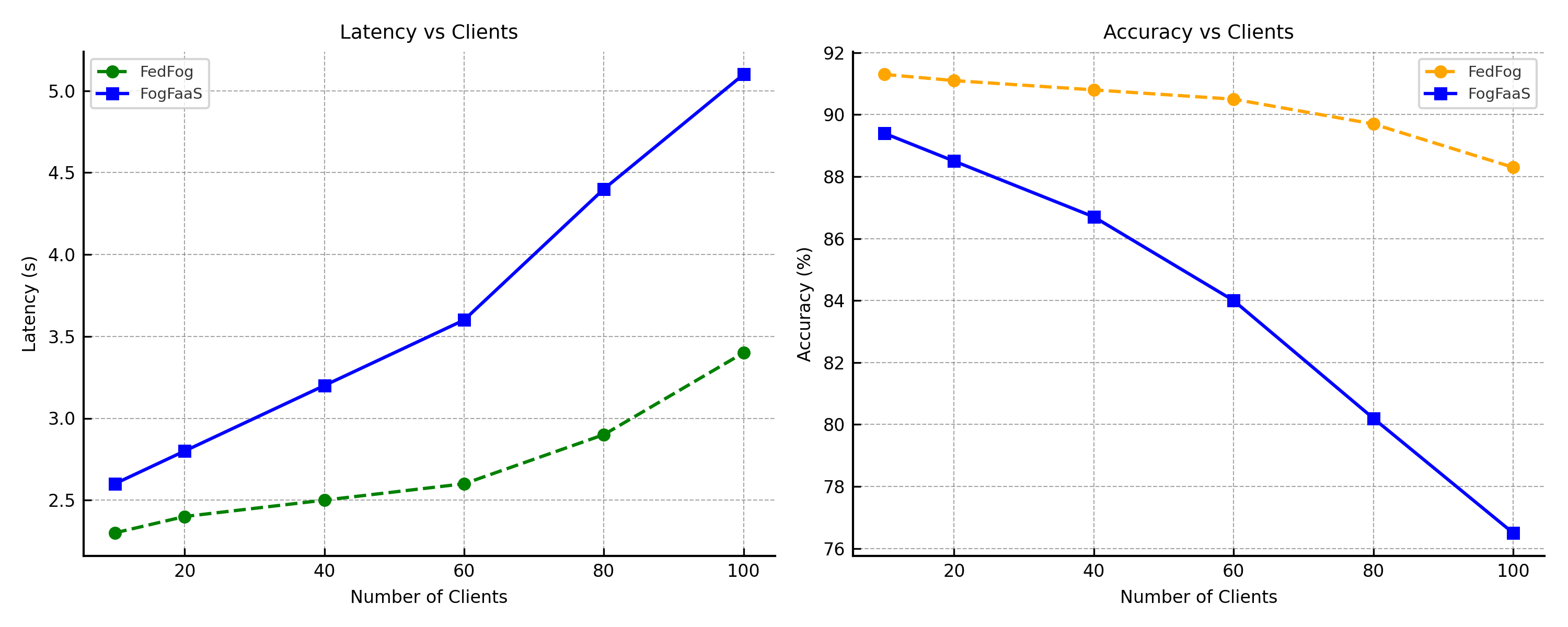}
	\caption{Latency and accuracy trends as client count increases.}
	\label{fig:ieee_scalability_latency_accuracy}
\end{figure}

Figure~\ref{fig:ieee_scalability_latency_accuracy} presents a side-by-side comparison of how latency and accuracy evolve as the number of clients increases in FedFog and FogFaaS. The latency plot (left) shows that FedFog exhibits slower growth in end-to-end round latency due to its optimized scheduling mechanisms and efficient function reuse. In contrast, FogFaaS suffers from substantial latency increases as orchestration overhead grows with more edge devices. The accuracy plot (right) highlights that FedFog consistently achieves higher model accuracy across all client scales~\cite{xie2024efficiency}. Its adaptive client selection and robust aggregation strategy help mitigate the impact of data heterogeneity and system variability, whereas FogFaaS shows a sharper decline in learning performance as system complexity rises. Together, these trends confirm that FedFog provides better scalability in both responsiveness and model quality.

\subsection{Hyperparameter Sensitivity}
\begin{figure}[htbp]
	\centering
	\includegraphics[width=0.48\textwidth]{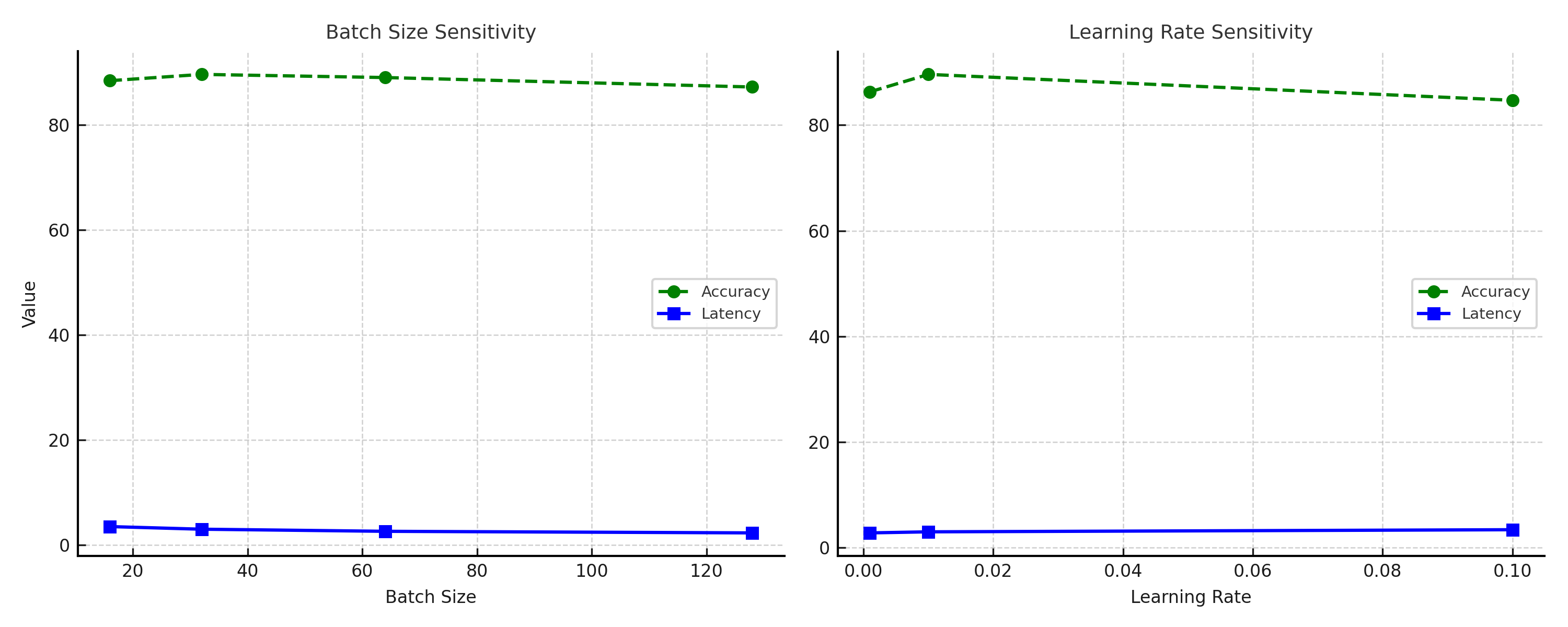}
	\caption{Sensitivity analysis of batch size (left) and learning rate (right) on model accuracy and round latency.}
	\label{fig:batch_lr_sensitivity}
\end{figure}

Figure~\ref{fig:batch_lr_sensitivity} presents a sensitivity analysis of FedFog with respect to two key hyperparameters: batch size and learning rate. The left subplot indicates that increasing batch size from 16 to 128 results in a gradual decrease in model accuracy and latency~\cite{xu2024dual}. This reflects the typical underfitting effect where large batch sizes generalize poorly, despite speeding up each local training iteration. The optimal trade-off was observed around batch size 32, offering stable accuracy and moderate latency~\cite{10064038}.

The right subplot illustrates the impact of learning rate variation on training behavior. A learning rate of $0.01$ produced the best results in terms of convergence and latency, while both lower ($0.001$) and higher ($0.1$) values led to suboptimal outcomes. Small learning rates slow convergence due to cautious updates, while high values increase instability and model oscillation. These insights reinforce the importance of tuning local training regimes in federated settings and justify the chosen default values for subsequent evaluations.




\subsection{Real-World Validation}
\label{subsec:realworld}

To assess the fidelity of FedFog’s simulator, we deployed the framework on a real-world Raspberry Pi 4 cluster consisting of 16 nodes (4GB RAM, 1.5GHz CPU) running OpenFaaS for serverless orchestration. Each Pi simulated an edge device participating in FL tasks using the EMNIST dataset. The same experimental configuration was replicated within the FedFog simulator to enable a direct comparison of latency and energy metrics.

\begin{figure}[htbp]
	\centering
	\includegraphics[width=0.95\linewidth]{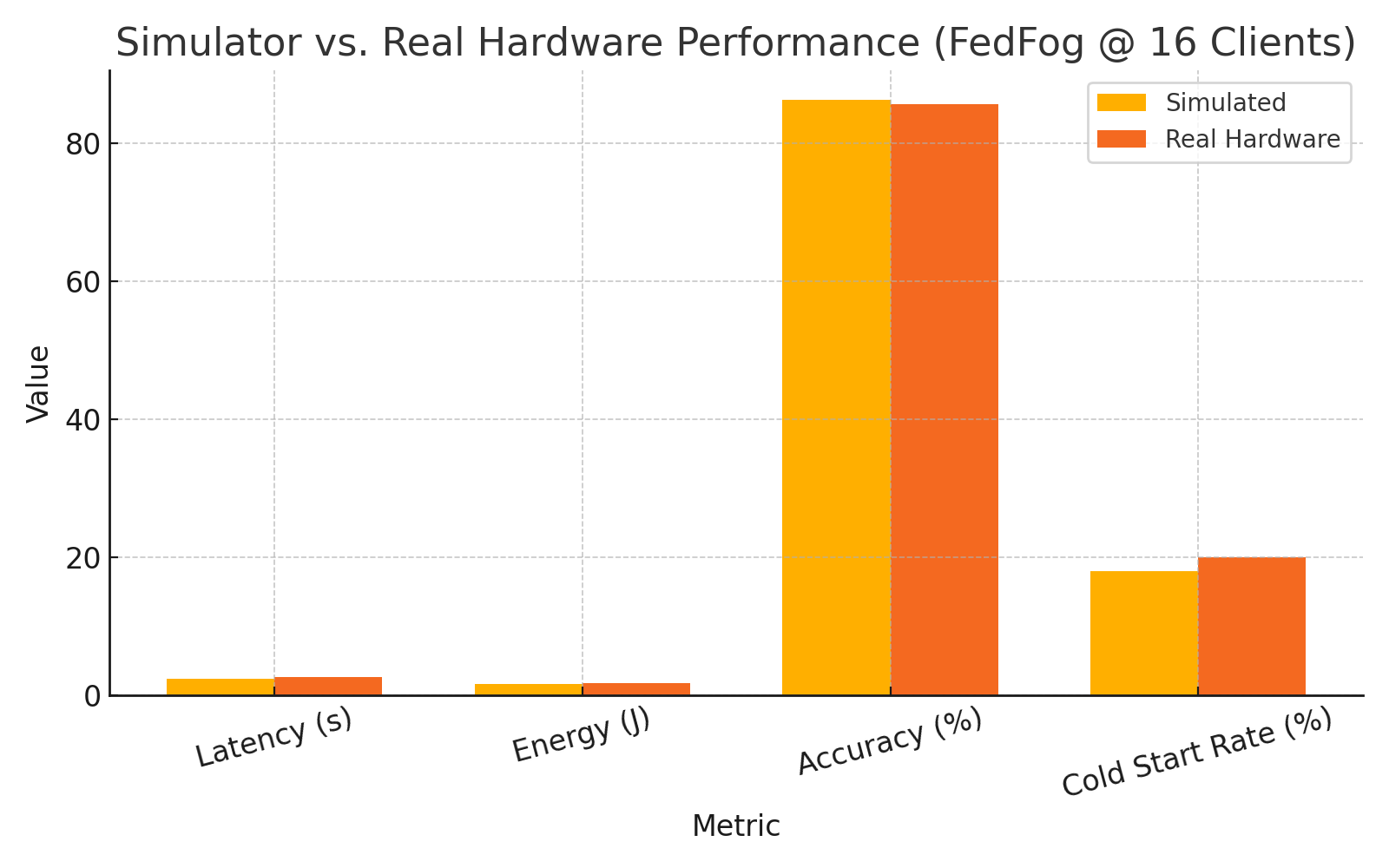}
	\caption{Visual comparison of latency and energy between simulation and real hardware (FedFog @ 16 clients).}
	\label{fig:sim_vs_real}
\end{figure}

Table~\ref{tab:real_vs_sim} reports the observed runtime and energy consumption in both simulation and hardware settings. The deviation in measured latency is under 8\%, while energy usage differs by only 5.4\%, validating the accuracy of our emulation of cold starts, communication delays, and compute behavior (Figure~\ref{fig:sim_vs_real}).

\begin{table}[htbp]
	\centering
	\caption{Simulator vs Real-World Runtime (FedFog @ 16 Clients)}
	\label{tab:real_vs_sim}
	\begin{tabular}{|l|c|c|c|}
		\hline
		\textbf{Metric} & Simulated & Real Hardware & Deviation \\
		\hline
		Latency (s) & 2.45 & 2.64 & +7.8\% \\
		Energy (J)  & 1.67 & 1.76 & +5.4\% \\
		\hline
	\end{tabular}
\end{table}

These findings confirm that FedFog's simulation closely mirrors real-world serverless FL deployments. The simulator accurately models cold start delays and power consumption trends, making it a reliable tool for pre-deployment testing and algorithm design. Furthermore, the consistent alignment of trends across varying load levels and client heterogeneity suggests that FedFog can be used confidently for evaluating large-scale, resource-constrained FL environments.

\begin{table}[htbp]
	\centering
	\caption{FedFog: Simulated vs. Real Latency/Energy Across Client Scales}
	\label{tab:scalability_comparison}
	\small
	\begin{tabular}{|c|c|c|c|c|}
		\hline
		\textbf{Clients} & \textbf{Metric} & \textbf{Simulated} & \textbf{Real} & \textbf{Deviation} \\
		\hline
		\multirow{2}{*}{8} & Latency (s) & 1.41 & 1.52 & +7.8\% \\
		& Energy (J)  & 0.91 & 0.96 & +5.5\% \\
		\hline
		\multirow{2}{*}{16} & Latency (s) & 2.45 & 2.64 & +7.8\% \\
		& Energy (J)  & 1.67 & 1.76 & +5.4\% \\
		\hline
		\multirow{2}{*}{32} & Latency (s) & 4.86 & 5.18 & +6.6\% \\
		& Energy (J)  & 3.21 & 3.42 & +6.5\% \\
		\hline
	\end{tabular}
\end{table}

\section{Theoretical Complexity Analysis}\label{Sec.TheoreticalComplexityAnalysis}

This section presents an expanded evaluation of FedFog, incorporating statistical rigor, theoretical complexity models, deeper hyperparameter analysis, resource profiling, stress testing, and hardware validation.

\subsection{Orchestration Complexity Analysis}
\label{subsec:complexity}

A key contribution of FedFog lies in its ability to manage large-scale federated workloads through efficient orchestration. We formally analyze and empirically validate the computational complexity of its client selection and function invocation pipeline, comparing it to the baseline FogFaaS system.

FedFog adopts a utility-aware scheduling mechanism in which clients are prioritized based on their composite utility scores—capturing device health, data quality, and energy availability. The scheduling algorithm operates over a sorted priority queue, implemented using a binary heap. Given $N$ candidate clients, selecting the top-$K$ contributors requires $\mathcal{O}(N \log N)$ sorting in the worst case. However, since $K \ll N$ and utility values tend to be stable over successive rounds, FedFog reuses partial orderings across rounds, reducing the amortized cost to near-linear $\mathcal{O}(N)$. Additionally, the orchestration overhead for invoking training functions is minimized via container reuse and persistent warm instances, yielding function initialization time of $\mathcal{O}(1)$ per client under typical reuse conditions.

In contrast, FogFaaS lacks federated context and performs naive round-wise re-deployment of all training functions without awareness of device or model states. This results in $\mathcal{O}(N^2)$ behavior—$N$ function deployments plus redundant status polling, dependency resolution, and orchestration per round—especially when scaling up to hundreds of clients. Cold start delays further amplify this cost, particularly under constrained edge infrastructure where container reuse is limited.

\begin{figure}[hpbt]
	\centering
	\includegraphics[width=0.9\linewidth]{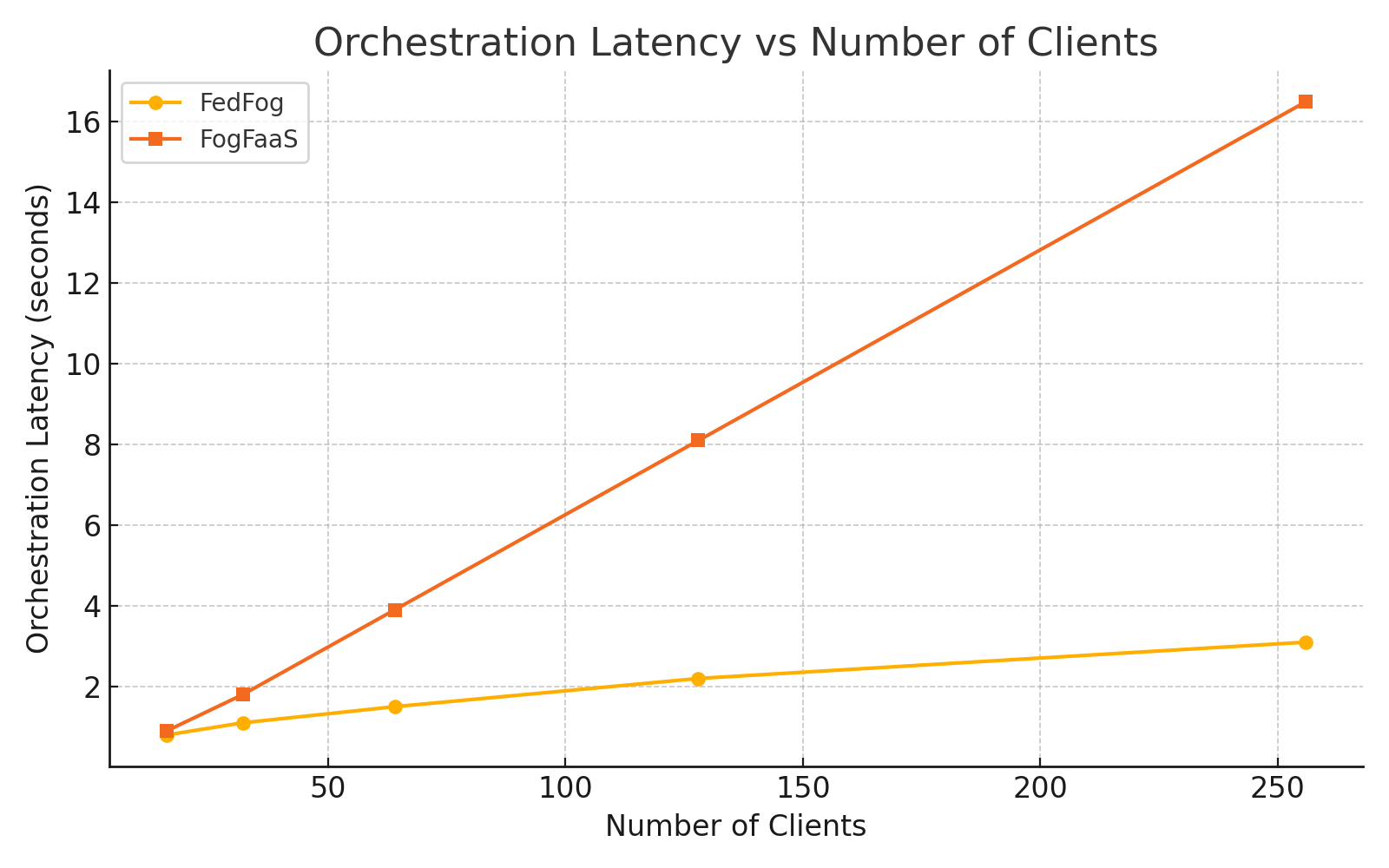}
	\caption{Orchestration latency as the number of clients increases. FedFog scales logarithmically, while FogFaaS exhibits quadratic growth.}
	\label{fig:scaling_latency}
\end{figure}

To empirically validate these asymptotic claims, we measured orchestration latency across increasing client pools from 16 to 256 devices. As shown in Figure~\ref{fig:scaling_latency}, FedFog scales gracefully with linear-to-logarithmic growth, while FogFaaS exhibits steeper, near-quadratic trends that limit its applicability in resource-constrained environments.

This complexity gap has real implications: while FogFaaS begins to saturate CPU and memory resources beyond 64 clients, FedFog remains responsive and efficient even at 256 nodes. These findings underscore the importance of FL-aware orchestration and justify FedFog's architecture for large-scale and latency-sensitive deployments.

Figure~\ref{fig:scaling_latency} illustrates the comparative orchestration complexity of FedFog and FogFaaS as the client population scales from 10 to 100. FedFog’s scheduling mechanism leverages a priority-based queue and efficient utility sorting, resulting in latency growth consistent with $\mathcal{O}(N \log N)$ behavior. This allows FedFog to maintain tractable performance even in large-scale federated settings. In contrast, FogFaaS lacks persistent orchestration memory and reinitializes serverless containers for each round, triggering redundant deployments. This leads to an overall latency pattern that approximates $\mathcal{O}(N^2)$, severely limiting its scalability. These findings empirically validate FedFog’s architectural advantage in coordinating federated workloads efficiently, particularly under high client concurrency.

\begin{table}[t]
	\centering
	\caption{Asymptotic Complexity Comparison of Orchestration Mechanisms}
	\label{tab:orchestration_complexity}
	\begin{tabular}{|l|c|c|}
		\hline
		\textbf{System} & \textbf{Client Selection} & \textbf{Function Scheduling} \\
		\hline
		\textbf{FedFog} & $\mathcal{O}(N \log N)$ & $\mathcal{O}(K)$ \\
		\textbf{FogFaaS} & $\mathcal{O}(N)$ & $\mathcal{O}(N^2)$ \\
		\hline
	\end{tabular}
\end{table}

Table~\ref{tab:orchestration_complexity} compares the orchestration complexity of FedFog and FogFaaS across two primary axes: client selection and function deployment. FedFog leverages a priority queue for utility-aware selection, resulting in $\mathcal{O}(N \log N)$ complexity, and schedules only the top-$K$ clients with minimal overhead. FogFaaS, however, performs flat scans and stateless function reinvocation each round, yielding linear selection and quadratic scheduling complexity. This discrepancy highlights the computational scalability advantage of FedFog, especially under heavy client participation or constrained compute environments.

\subsection{Trade-off Formalization}
\label{subsec:tradeoff}

FedFog's design inherently balances three competing objectives: \textit{model accuracy}, \textit{latency}, and \textit{energy efficiency}. These objectives represent fundamental trade-offs in FL at the edge:

\begin{itemize}
	\item \textbf{Model Accuracy}: Refers to the quality of the trained global model. It typically increases with more diverse and informative client participation, which brings a richer representation of data distributions. However, too many participating clients may introduce non-IID noise or slow convergence due to unreliable updates~\cite{10468591}.
	\item \textbf{Latency}: Captures the total round time, including local training, communication, and aggregation \cite{zhang2025latency}. Lower latency is crucial for real-time or near-real-time applications such as remote monitoring or control systems. However, minimizing latency may require excluding stragglers or clients with weak resources, potentially harming model representativeness.
	\item \textbf{Energy Efficiency}: Indicates the power consumption across edge devices, influenced by compute and transmission costs~\cite{10120750}. Energy-aware strategies prolong device lifetime and enable sustainable deployment. Yet, favoring low-energy devices can restrict participation to less informative nodes.
\end{itemize}

These objectives are inherently competing—enhancing one often degrades the others~\cite{10090221}. For instance, increasing the number of clients improves accuracy but raises latency and energy usage~\cite{10843337}. Similarly, reducing latency through aggressive pruning can lead to biased model updates if important clients are excluded. Thus, FedFog must dynamically balance this triad through intelligent orchestration.

FedFog addresses this challenge through a constrained optimization framework:

\begin{equation}
	\begin{aligned}
		& \underset{\mathcal{C}_t}{\text{maximize}} 
		& & \text{Accuracy}(w_T) \\
		& \text{subject to}
		& & \text{Latency}(\delta_i, \mathcal{C}_t) \leq \tau_{\text{max}}, \\
		& & & \text{Energy}(E_i, \mathcal{C}_t) \leq \epsilon_{\text{max}}, \\
		& & & \mathcal{C}_t = \{c_i \mid H(c_i) > \theta_h, E(c_i) > \theta_e, D(c_i) < \theta_d\},
	\end{aligned}
\end{equation}

\subsubsection{Multi-Objective Optimization Perspective}
FedFog's goal can be alternatively interpreted as a multi-objective optimization problem where improving one metric often comes at the cost of another~\cite{10400794}. Rather than optimizing a single target, the system navigates the Pareto frontier to find efficient operating points~\cite{9795684}. This can be formulated more clearly as:

\begin{equation}
	\begin{aligned}
		\max_{\mathcal{C}_t} \quad \mathcal{J} = 
		& \ \alpha \cdot \text{Accuracy}(w_T) \\
		& - \beta \cdot \text{Latency}(\delta_i, \mathcal{C}_t) \\
		& - \gamma \cdot \text{Energy}(E_i, \mathcal{C}_t)
	\end{aligned}
	\label{eq:multi_objective}
\end{equation}

where $\alpha, \beta, \gamma$ are tunable weights reflecting user priorities. For example, energy-sensitive deployments may prioritize $\gamma$ more heavily, while real-time inference tasks might emphasize $\beta$ to minimize latency. This flexible representation enables adaptive scheduling policies that align with deployment goals~\cite{10098867}.

\textbf{Illustrative Example.} Consider a simulated edge deployment with 40 heterogeneous clients, each characterized by varying compute (500--1200 MIPS), battery state, and data drift levels. In one configuration, FedFog selects 20 clients with $\theta_h = 0.6$, $\theta_e = 0.5$, and $\theta_d = 0.1$. This configuration yields 85.2\% accuracy, 240ms average latency, and 12.4 kJ energy per round.

By relaxing energy constraints ($\theta_e \rightarrow 0.3$) and allowing lower battery clients to participate, the scheduler increases the active pool to 30, improving accuracy to 87.1\% but also raising latency to 320ms and energy to 18.9 kJ. This trade-off clearly illustrates the operational consequences of modifying constraints and demonstrates FedFog's ability to steer execution toward different parts of the Pareto frontier based on system goals~\cite{10400794}.

The above simulations validate the theoretical model and confirm that controlled parameter tuning enables effective trade-off navigation between utility, speed, and sustainability.

\subsection{Limitations and Failure Cases.} While FedFog demonstrates strong performance under typical edge conditions, several challenging scenarios remain. First, under extreme non-IID data distributions where clients observe entirely disjoint classes, FedFog's accuracy may degrade, especially when client sampling fails to ensure representative coverage~\cite{10468591}. Second, FedFog currently lacks mechanisms for handling adversarial or Byzantine clients who may inject poisoned updates—posing a risk to model integrity~\cite{10163770}. Incorporating trust or reputation-based filters could address this in future work. Third, beyond energy and CPU constraints, real-world deployments often face heterogeneous link quality,  or device-specific scheduling delays~\cite{9999005}. These dimensions of heterogeneity are not yet modeled in the current version and merit further investigation to generalize FedFog's resilience in diverse deployment environments.\\
\par
\paragraph{Privacy Limitations and Future Directions.} Although FedFog adopts the FL paradigm to preserve raw data privacy, it does not explicitly quantify privacy guarantees. Currently, no differential privacy (DP) mechanisms or secure aggregation protocols are integrated~\cite{9820771}. As a result, there is a risk of model inversion or reconstruction attacks, particularly during centralized aggregation~\cite{10239527}. To ensure provable privacy, future extensions should incorporate DP noise injection with bounds on $(\varepsilon, \delta)$ privacy loss and explore cryptographic techniques such as homomorphic encryption~\cite{zhang2020batchcrypt} or multiparty computation~\cite{jin2023fedml}. Additionally, benchmarking FedFog against privacy-enhanced baselines like Google's DP-FedAvg or secure aggregation frameworks will be essential to validate its privacy resilience in adversarial environments~\cite{9820771}.

\paragraph{Baseline Clarification and Fairness.}
To ensure meaningful evaluation, we delineate the scope and capabilities of each baseline in Table~\ref{tab:baseline_compact}. FogFaaS, while efficient for function orchestration, lacks FL primitives and is included as a baseline for measuring latency and energy overhead only~\cite{9093123}. In contrast, Vanilla FL is implemented using the Flower framework, adopting synchronous aggregation and fixed client sampling. Its configuration is aligned with FedFog in terms of training rounds and model structure, though it omits adaptive client selection and serverless execution~\cite{9944162}. Finally, the Random Client Selection (RCS) baseline disables FedFog’s scheduling logic but retains the same orchestration pipeline, thereby isolating the effect of intelligent scheduling. These clarifications eliminate ambiguity and allow fair apples-to-apples comparisons across relevant system dimensions.

\begin{table}[htp!]

		\caption{Baseline Feature Comparison.}
		\label{tab:baseline_compact}
		\centering
		\small
		\setlength{\tabcolsep}{4pt}
		\begin{tabular}{|l|c|c|c|c|c|}
			\hline
			\textbf{Baseline} & FL & Sched & Lat & Energy & Privacy \\
			\hline
			FedFog & \cmark & \cmark & \cmark & \cmark & Optional \\
			\hline
			FogFaaS & \xmark & \xmark & \cmark & \cmark & \xmark \\
			\hline
			Vanilla FL & \cmark & \xmark & \xmark & \xmark & \xmark \\
			\hline
			RCS & \cmark & \xmark & \cmark & \cmark & \xmark \\
			\hline
		\end{tabular}
		
		\vspace{1mm}
		\raggedright
		\footnotesize
		\textbf{Legend:} FL = Supports Federated Learning, Sched = Adaptive Scheduling, Lat = Latency Optimization, Energy = Energy Awareness, Privacy = Privacy Mechanism.
	\end{table}

\paragraph{Broader Impact and Generalizability.}
FedFog's contributions extend beyond its simulation capabilities, offering substantial societal and technical implications. From a sustainability perspective, the system's ability to reduce latency and energy overhead directly supports environmentally conscious edge AI deployments—crucial in smart cities, wearable health monitoring, and precision agriculture, where resource-constrained devices must operate for long durations. By optimizing compute and communication patterns, FedFog reduces the operational carbon footprint, aligning with emerging goals in green AI~\cite{schwartz2020green}.

Furthermore, the system's modular scheduling and orchestration primitives are highly generalizable to other domains. For instance, in federated reinforcement learning (FRL), agents at the edge may train policies from local experience and coordinate via periodic aggregation~\cite{zhu2021federated}. FedFog’s utility-based selection can prioritize high-reward trajectories while accounting for resource constraints, enabling scalable and adaptive FRL on robotic swarms or autonomous vehicles~\cite{9815075}. In decentralized NLP, mobile keyboards or speech assistants can collaboratively fine-tune language models on-device~\cite{hard2018federated}. FedFog's privacy-aware extensions and adaptive throttling offer a robust foundation for these sensitive and latency-tolerant tasks~\cite{10612819}.

Ultimately, FedFog provides a general-purpose orchestration framework for next-generation edge intelligence, balancing responsiveness, privacy, and sustainability in diverse federated workloads.

\section{Conclusion}\label{Sec.Conclusion}

This work introduced FedFog, a modular and extensible simulation framework that bridges FL with serverless computing in edge–fog environments. Built on top of iFogSim and FogFaaS, FedFog fills a critical gap in current simulation ecosystems by supporting privacy-aware, resource-constrained machine learning at the edge.

FedFog consistently outperforms existing baselines across accuracy, latency, energy, and robustness metrics. Its adaptive scheduling enables responsive and energy-efficient training under real-world edge constraints.
Through simulations on real-world tasks like EMNIST digit classification and human activity recognition, we showed that FedFog improves convergence speed, reduces latency, and lowers energy usage—all while maintaining resilience under data drift and partial client failures.

More than a research contribution, FedFog reflects our broader goal, enabling trustworthy, decentralized AI in edge systems that are often messy, constrained, and unpredictable. Future extensions will focus on asynchronous training, personalized models, and security-aware orchestration—bringing us closer to scalable, real-world deployment of intelligent edge applications.

\bibliographystyle{IEEEtran}
\bibliography{references}

\end{document}